 \documentclass[twocolumn,twocolappendix]{aastex631}

\shorttitle{Light Curve Model of V392 Per}
\shortauthors{Hachisu \& Kato}

\begin{document}

\title{A multiwavelength light curve analysis of the classical nova V392 Per:\\
Optical contribution from an irradiated accretion disk
during the nova wind phase}


\author[0000-0002-0884-7404]{Izumi Hachisu}
\affil{Department of Earth Science and Astronomy, 
College of Arts and Sciences, The University of Tokyo,
3-8-1 Komaba, Meguro-ku, Tokyo 153-8902, Japan} 
\email{izumi.hachisu@outlook.jp}


\author[0000-0002-8522-8033]{Mariko Kato}
\affil{Department of Astronomy, Keio University, 
Hiyoshi, Kouhoku-ku, Yokohama 223-8521, Japan} 

\begin{abstract}
The classical nova V392 Per 2018 is characterized by a very fast optical
decline, long binary orbital period of 3.23 days, 
detection of GeV gamma rays, and almost identical decay trends of
$B$, $V$, and $I_{\rm C}$ light curves.
The last feature is unique because most novae develop strong emission lines
in the nebular phase and these lines contribute especially to the $B$ and
$V$ bands and make large differences between the $BV$ and $I_{\rm C}$
light curves.  This unique feature can be understood if the optical flux 
is dominated by continuum until the late phase of the nova outburst.
Such continuum radiation is emitted by a bright accretion disk irradiated
by a hydrogen burning white dwarf (WD) and viscous heating disk with
high mass-accretion rate after the hydrogen burning ended.
We present a comprehensive nova outburst
model that reproduces all of these light curves.  
We determined the WD mass to be $M_{\rm WD}=1.35$ - $1.37 ~M_\sun$
and the distance modulus in the $V$ band to be $(m-M)_V=14.6 \pm 0.2$;
the distance is $d= 3.45\pm 0.5$ kpc for the reddening of $E(B-V)=0.62$.
\end{abstract}




\section{Introduction}
\label{introduction}
A classical nova is a thermonuclear explosion of a hydrogen-rich envelope
on a mass-accreting white dwarf (WD) \citep[][for a recent review]{del20i}.
Hydrogen ignites to trigger an outburst when the mass of the envelope
reaches a critical value \citep[e.g.,][for a recent fully self-consistent
nova explosion model]{kat22sha}.  V392 Per had been known as a dwarf nova
\citep[$V\sim 15$--$17$ in quiescence, e.g.,][]{dow01ws}
before the nova outburst in 2018,
where dwarf novae are much fainter phenomena than classical novae and their
outbursts are triggered by thermal instability of an accretion disk
\citep[e.g.,][for a review]{osa96}. 

A 10 mag brightening of V392 Per was discovered at 6.2 mag by Y. Nakamura
on UT 2018 April 29.474 ($=$JD 2,458,237.974, cf. CBET 4515).
%
Follow-up spectroscopy by \citet{wag18td} confirmed that it is
a classical nova.  Immediately after the discovery, it was well
observed in multiwavelength bands, especially in optical 
\citep{mun20mm, cho21sh}
as well as gamma-ray \citep{alb22aa} and X-ray \citep{mur22dh}. 

\begin{figure}
\gridline{\fig{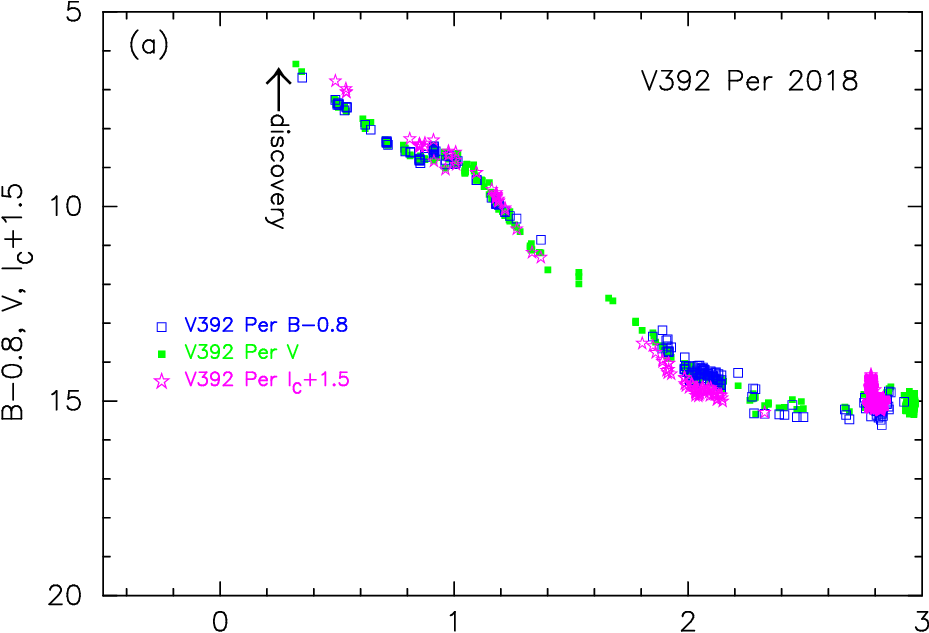}{0.45\textwidth}{}
          }
\gridline{\fig{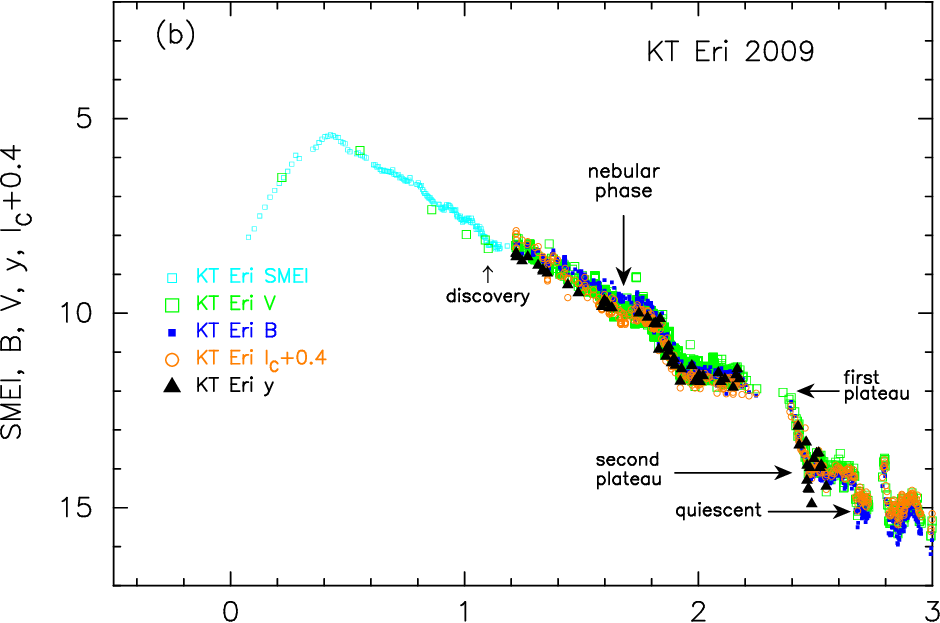}{0.45\textwidth}{}
          }
\gridline{\fig{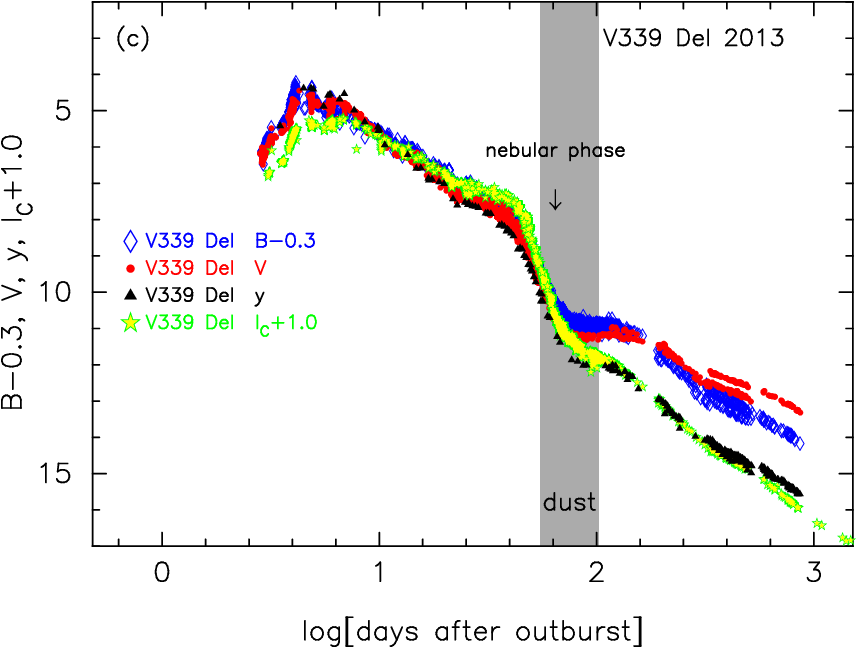}{0.45\textwidth}{}
          }
\caption{
(a) The $B,V,I_{\rm C}$ light curves of the 2018 outburst
of V392 Per against a logarithmic time.  We assume that the outburst day is
$t_{\rm OB}=$JD 2,458,236.2 ($=$UT 2018 April 27.7).
(b) The 2009 outburst of KT Eri. 
The data are the same as those in \citet{hac25kw}.
(c) The 2013 outburst of V339 Del.  
The data are the same as those in \citet{hac24km}.
\label{V392_per_kt_eri_v339_del_bvi_logscale}}
\end{figure}

V392 Per is characterized by 
(1) GeV gamma-rays detected from just after discovery to 7 days later, and
(2) fast decline time by 2 or 3 mag from maximum, i.e.,
$t_2=3$ or $t_3=11$ days \citep{mun20mm},
which suggests a very massive WD close to
the Chandrasekhar mass limit.
(3) Late $V$ magnitude almost saturated at $V\sim 15.3$,
but it is about 2 mag brighter than the preoutburst brightness
\citep{mun20mm}.
(4) Post-nova spectrum-energy-distributions (SEDs) indicate high mass
transfer rate, which could be driven by irradiation from the WD
\citep{mur22dh}.  
(5) The decline trends of $BVI_{\rm C}$ light curves are almost overlapped
during the outburst from near maximum to quiescence as shown in Figure
\ref{V392_per_kt_eri_v339_del_bvi_logscale}a.  

The last (5th) feature was clearly identified in the classical nova KT Eri
\citep[Figure \ref{V392_per_kt_eri_v339_del_bvi_logscale}b; ][]{hac25kw},
while typical novae show different decline trends in the $BVI_{\rm C}$
bands \citep[see V339 Del in Figure 
\ref{V392_per_kt_eri_v339_del_bvi_logscale}c, ][for an example]{hac24km}. 

In V392 Per and KT Eri, both the colors of $B-V$ and $V-I$ vary very little
during the nova outburst.  This feature can be confirmed
in the color-magnitude diagrams $(B-V)_0$-$M_V$ and $(V-I)_0$-$M_I$ as shown
in Figure \ref{hr_diagram_v392_per_kt_eri_bv_vi}.
The tracks go down almost straight in both the color-magnitude diagrams
while the other novae traverse largely from left to right (or
right to left) like in LV Vul (orange line) and V1500 Cyg (green line).
See Figure 12 of \citet{hac25kw} for the color-magnitude diagrams for
V339 Del, which clearly show that V339 Del belongs to the traverse type
of tracks like LV Vul.

\citet{hac16kb} 
showed the $(B-V)_0$-$M_V$ color-magnitude diagrams
for the total of 42 novae. Among them, 23 show a traverse type such as
LV Vul (orange lines in Figure \ref{hr_diagram_v392_per_kt_eri_bv_vi}),
only one (U Sco) shows a straight-down
type such as KT Eri.  The other 10 novae are too short to determine
the type and 8 novae are also too short by dust blackout.
\citet{hac21k} 
showed both the $(B-V)_0$-$M_V$ and
$(V-I)_0$-$M_I$ color-magnitude diagrams for the total of 53 novae.
Among them, 39 show
a traverse type while only one (U Sco) shows a straight-down type.
The tracks of the other 5 novae are too short to determine the type.
The other 3 novae experienced dust black-out and their tracks are too short.
The remaining 5 novae have a red giant companion and the colors are dominated
by the red giant companion in the later phase of a nova outburst.
Thus, the 5th feature of V392 Per is quite rare among many classical novae.

\citet{hac25kw} explained the almost overlapping trend of $BVyI_{\rm C}$
light curves in KT Eri when continuum flux dominates
line fluxes during the outburst.  They reproduced the $BVyI_{\rm C}$ 
light curves by calculating the summation of the free-free
emission from the nova wind and each photospheric emission
from the binary components (WD, accretion disk, and companion star).


In the present paper, we try to give theoretical explanations on
these properties (1)--(5) based on our nova model \citep{hac25kw}
and to determine the physical properties of V392 Per
such as the WD mass and mass-accretion rate to the WD.

Our paper is organized as follows.  First we summarize the observational
results and then give a short interpretation to them
in Section \ref{observation_summary_interpretation}.
Our shock model and expected emission are presented and discussed in Section
\ref{discussion}.  Conclusions follow in Section \ref{conclusions}.
Appendix gives model details (Appendix \ref{optically_thick_wind_model}),
various methods for obtaining distance modulus, extinction,
and distance to a nova,
as well as the time-stretching method for nova light curves
(Appendix \ref{time_stretching_method}).

\section{Observational summary and quick interpretation}
\label{observation_summary_interpretation}

Before showing our model light curve fitting, we list the physical
properties of V392 Per that we must take into account from the
theoretical points of view.
Figure \ref{V392_per_kt_eri_v339_del_bvi_logscale} shows
the optical/NIR $BVI_{\rm C}$ light curves of V392 Per, KT Eri, and
V339 Del.  The data of $BVI_{\rm C}$ of V392 Per are taken from
the archives of the American Association of Variable Star Observers
(AAVSO), the Variable Star Observers League of Japan (VSOLJ), and
\citet{mun20mm}.  For comparison, in Figure 
\ref{V392_per_kt_eri_v339_del_bvi_logscale}b and
\ref{V392_per_kt_eri_v339_del_bvi_logscale}c,
we added Str\"omgren $y$ magnitudes to the KT Eri and V339 Del light curves,
the data of which are the same as those in
\citet{hac25kw} and \citet{hac24km}, respectively.

\begin{figure*}
\gridline{\fig{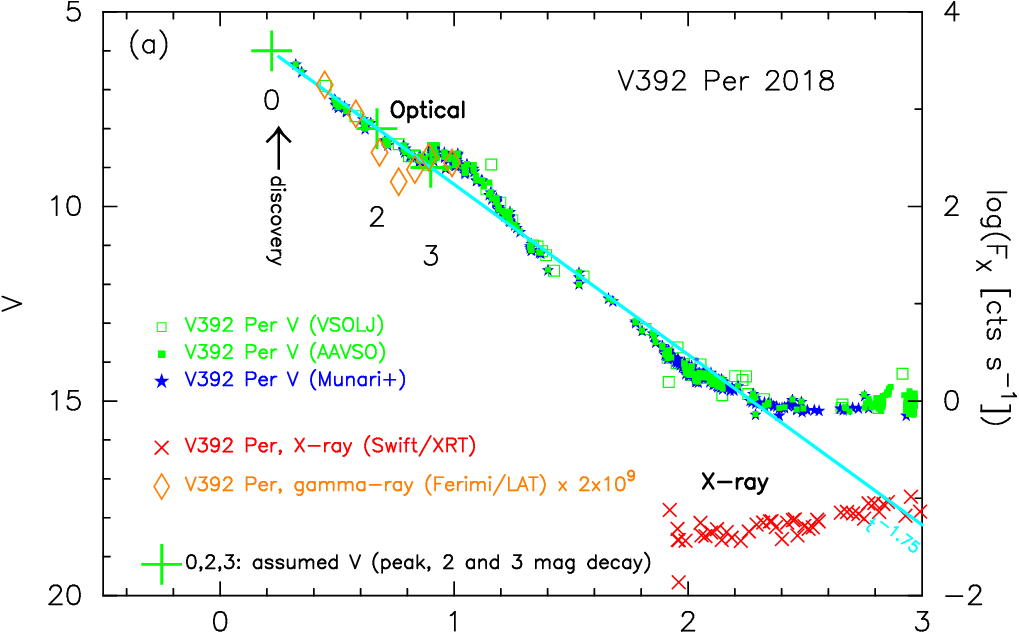}{0.7\textwidth}{}
          }
\gridline{\fig{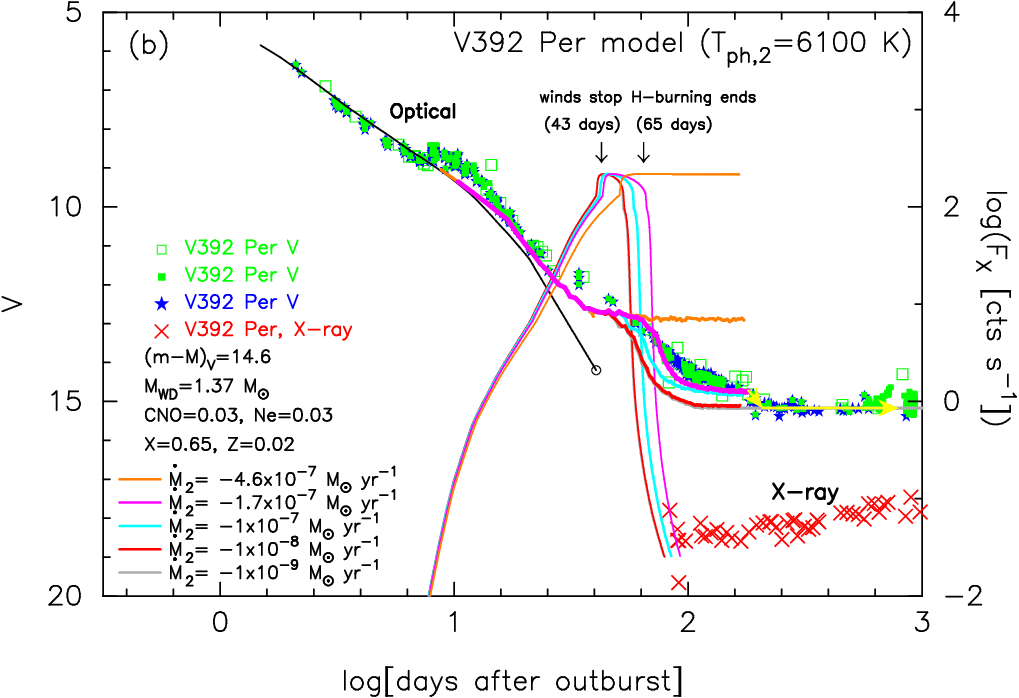}{0.7\textwidth}{}
          }
\caption{
(a) The $V$, X-ray, and gamma-ray light curves of the 2018 outburst
of V392 Per against a logarithmic time.  We assume that the outburst day is
$t_{\rm OB}=$JD 2,458,236.2 ($=$UT 2018 April 27.7).
The $V$ data are taken from the Variable Star Observers League of Japan
(VSOLJ), the American Association of Variable Star Observers (AAVSO),
and \citet{mun20mm}.
The Swift X-ray (0.3--10.0 keV) count rates are also added,
taken from the Swift website \citep{eva09}.
We also add the 0.3--100 GeV gamma-ray flux detected with the Fermi/LAT
\citep{alb22aa}.  The large green plus (+) symbols indicate the assumed
0 mag (peak, labeled 0), 2 mag (2), and 3 mag (3) decay from the peak.  
(b) Same as panel (a), but we overplot our model light curves 
for the distance modulus in the $V$ band of $(m-M)_V=14.6$.
We assume a Roche-lobe-filling companion star with the photospheric
temperature of $T_{\rm ph, 2}= 6,100$ K and mass of $M_2= 1.0 ~M_\sun$.
The black line is our free-free (FF) + photospheric blackbody (BB)
model light curve of a $1.37 ~M_\sun$ white dwarf (WD).
Including the brightnesses of the disk and companion star irradiated by
the hot WD and viscous heating of the disk, we plot the thick orange,
magenta, cyan, red, and gray lines of our model $V$ light curves for the
$1.37 ~M_\sun$ WD (Ne3) with different mass-accretion rates in
Table \ref{post-outburst_parameters}.  The thin colored lines
are for the corresponding model X-ray fluxes (0.3--10.0 keV),
although the thin red and gray lines are overlapped. The yellow arrows
indicate the path in which we reduce $-\dot{M}_2= 1.7\times 10^{-7}$
to $1.0\times 10^{-9} ~M_\sun$ yr$^{-1}$.
\label{v392_per_only_v_x_big_disk_6100k_logscale}}
\end{figure*}

Figure \ref{v392_per_only_v_x_big_disk_6100k_logscale}a shows 
optical ($V$), X-ray, and gamma-ray light curves of the 2018 outburst
of the classical nova V392 Per.  
The gamma-ray data are from \citet{alb22aa} and the X-ray data are
from the Swift website \citep{eva09}. 

\begin{figure*}
\gridline{\fig{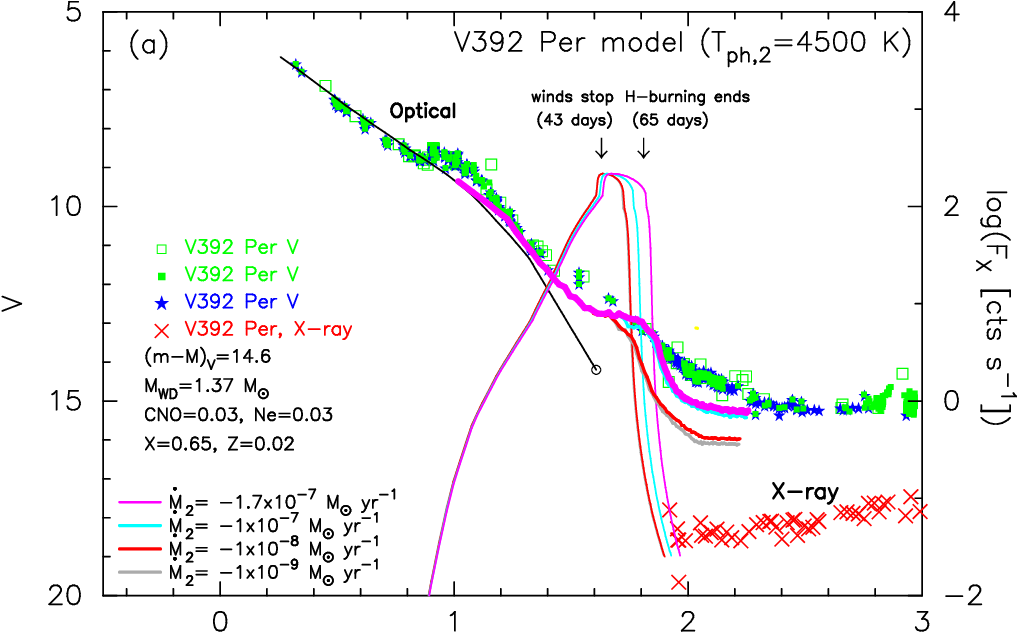}{0.7\textwidth}{}
          }
\gridline{
          \fig{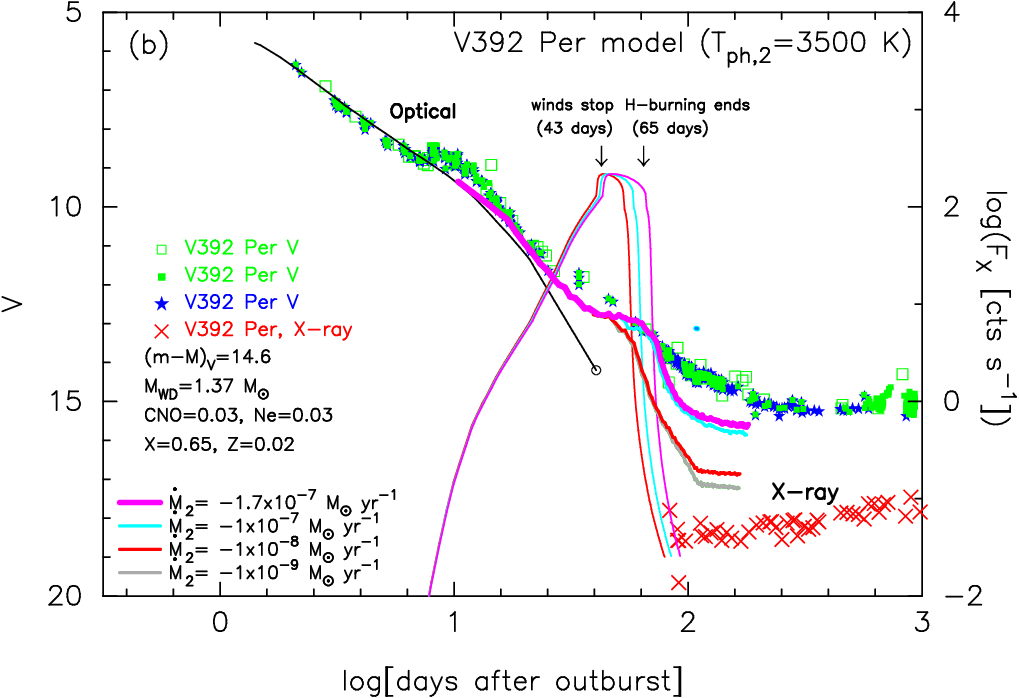}{0.7\textwidth}{}
          }
\caption{
Same as those in Figure \ref{v392_per_only_v_x_big_disk_6100k_logscale}b,
but (a) for the photospheric temperature of the companion star,
$T_{\rm ph, 2}= 4,500$ K or (b) $T_{\rm ph, 2}= 3,500$ K.
The other symbols are the same as those
in Figure \ref{v392_per_only_v_x_big_disk_6100k_logscale}b.
\label{v392_per_only_v_x_big_disk_3500k_logscale}}
\end{figure*}

For later use and discussion, our model light curves of optical $V$ 
and supersoft X-ray are overplotted to the observational data in
Figures \ref{v392_per_only_v_x_big_disk_6100k_logscale}b,  
\ref{v392_per_only_v_x_big_disk_3500k_logscale}a, and 
\ref{v392_per_only_v_x_big_disk_3500k_logscale}b.
These model parameters are 
listed in Table \ref{post-outburst_parameters} and explained step by step.


\subsection{Distance, reddening, and orbital period}
\label{distance_reddening_orbit}

The distance to V392 Per is estimated by the Gaia eDR3 parallax to be
$d=3.45^{+0.62}_{-0.51}$ kpc \citep{bai21rf}.
We adopt the distance of $d= 3.45$ kpc in our model light curves. 
The reddening toward V392 Per was obtained by \citet{mun20mm} to be 
$E(B-V)= 0.72\pm 0.06$ from the intrinsic colors of $(B-V)_0= 0.23\pm 0.06$
near maximum and $(B-V)_0= -0.02\pm 0.04$ at $t_2$ \citep{van87y}.
The Bayestar2019 3D map of Galactic extinction \citep{gre19} reports
$E(B-V)= 0.62^{+0.03}_{-0.02}$ at the distance of $d=3.45$ kpc 
(see also Figure \ref{distance_reddening_relation_v392_per}
in Appendix \ref{distance-reddening_relation_v392_per}).

A pair of the distance and reddening can be constrained
by the relation \citep[e.g.,][]{rie85} of
\begin{equation}
(m-M)_V = 5 \log(d/{\rm 10~pc}) + 3.1 E(B-V).
\label{distance-reddening_law}
\end{equation}
The distance modulus in the $V$ band is obtained to be 
$\mu_V\equiv (m-M)_V= 14.6\pm 0.2$ toward V392 Per by comparing
its $V$ light curve with other well studied novae (see Appendix 
\ref{distance-reddening_relation_v392_per} for the time-stretching
method).  Inserting the distance of $d=3.45^{+0.62}_{-0.51}$ kpc and
the reddening of $E(B-V)= 0.62^{+0.03}_{-0.02}$ into 
Equation (\ref{distance-reddening_law}),
we also obtain the distance modulus
in the $V$ band of $(m-M)_V= 14.6\pm 0.4$, which is consistent with
the results in Appendix \ref{distance-reddening_relation_v392_per}
based on the time-stretching method.
In what follows, we use
$(m-M)_V= 14.6$, $d=3.45$ kpc, and $E(B-V)= 0.62$ unless otherwise specified.

The orbital period was first proposed by \citet{mun20mm}
to be $P_{\rm orb}= 3.4118\pm 0.0013$ days based on their $VRI$ photometry
post-eruption. \citet{schaefer22a} obtained $P_{\rm orb}= 3.21997\pm 
0.00039$ days based on the data of AAVSO and TESS.
\citet{mur22dh} presented a slightly different orbital period of
$P_{\rm orb}= 3.230\pm 0.003$ days with a caution on the
possible contamination by a nearby star 
($9\arcsec$ apart from V392 Per) in the AAVSO and TESS data.     
Therefore, we adopt $P_{\rm orb}= 3.230$ days.     
This period is relatively long among the classical novae with known 
orbital periods \citep{schaefer22a}.

\subsection{Optical peak and outburst day}
\label{photosphere_peak_outburstday}

Unfortunately, there are no data in the rising phase and around the 
optical maximum.  So, we do not know when the outburst (thermonuclear
runaway) occurs and when the optical brightness reaches its maximum.
The optical $V$ maximum probably occurs slightly before the
discovery day (JD 2,458,237.974).  \citet{mur22dh} discussed the property
of the $V$ light curve shape of V392 Per based on the morphology analysis
of \citet{str10sh}.  The shape of the V392 Per optical light curve belongs
to the P-class morphology.  In all the 19 P-class novae, their $t_3$ day
occur before the plateau phase.  If we assume that the $t_3$ day
occurred before the $V$ plateau phase starting from JD 2,458,244.14
(green plus mark labeled 3 in Figure 
\ref{v392_per_only_v_x_big_disk_6100k_logscale}a),
the $V$ brightness reaches maximum somewhat before the discovery. 
\citet{mur22dh} estimated the outburst day to be $t_{\rm OB}=$ JD 2,458,236.5
($=$ MJD 58236.0) and the $V$ maximum day $t_0=$ JD 2,458,237.6,
1.1 days after outburst, along their broken power-laws fitted with the
$V$ light curve.  They obtained $t_2=2.0$ day, $t_3=4.2$ day, and
$V_{\rm max}= 5.5$, $M_{V, \rm max}= -9.4$ for the
Gaia eDR3 parallax \citep[$d=3.5$ kpc,][]{bai21rf},
or $M_{V, \rm max}= -8.8$ for the MMRD relation \citep{del20i} and
$d=2.7$ kpc, both with $E(B-V)=0.7$ \citep{mun20mm}.

Here, we take a similar to, but a slightly different from,
\citet{mur22dh}'s way.
We assume that the $t_3$ day is located just before the $V$ plateau
phase begins on JD 2,458,244.14 at the green plus mark 
labeled 3 in Figure \ref{v392_per_only_v_x_big_disk_6100k_logscale}a.
To obtain the $V$ maximum day and brightness, we assume that the $V$ light
curve follows the universal decline law \citep{hac06kb} of 
\begin{equation}
L_V\propto t^{-1.75}
\label{universal_decline_law}
\end{equation}
from the $V$ maximum on day $t_0$ to the 3 mag decay day on JD 2,458,244.14,
where $L_V$ is the flux of
the $V$ band and $t$ is the time from the outburst ($t_{\rm OB}$),
as plotted by the thick cyan line in Figure 
\ref{v392_per_only_v_x_big_disk_6100k_logscale}a.
Assuming that the $t_3$ (3 mag decay) day occurred
just when the plateau phase started
on JD 2,458,244.14 (at the green plus mark labeled 3),
we obtain the $V$ maximum ($t_0$) day on $t_0 =$JD 2,458,237.86
(peak brightness, green plus mark labeled 0) 
and the $t_2$ day on JD 2,458,240.88 (2 mag below maximum, green plus mark
labeled 2), that is, $t_2=3.0$ days, $t_3= 6.3$ days,
$V_{\rm max}=6.0$.  Then, the maximum absolute brightness is 
$M_{V, \rm max}= -8.6$ for $(m-M)_V=14.6$ (see Section 
\ref{distance_reddening_orbit} for our recommended $(m-M)_V$).
The outburst day is obtained to be $t_{\rm OB}=$ JD 2,458,236.2 from the
time-stretching fit of V392 Per with \citet{kat22sha}'s self-consistent nova
outburst model (See Figure \ref{v392_per_kt_eri_v339_del_v_x_stretching}a
in Appendix \ref{outburst_day_from_time-streching}).
Then, the rising time to the peak
is about $(\Delta t)_{\rm rise}=10^{0.22}=$ 1.66 days.


\begin{deluxetable}{lllllll}
\tabletypesize{\scriptsize}
\tablecaption{Parameters and brightnesses of the post-outburst phase
\label{post-outburst_parameters}}
\tablewidth{0pt}
\tablehead{
\colhead{$M_{\rm WD}$} & \colhead{$T_{\rm ph,2}$} & \colhead{$-\dot{M}_2$}
& \colhead{$\alpha$} & \colhead{$\beta$} & \colhead{$V$\tablenotemark{a}}
& comment \\
($M_\sun$) & (~K~)  & ($M_\sun$ yr$^{-1}$) &  &  &  ( mag ) &
}
\startdata
1.37 & 6100 & $4.6\times 10^{-7}$ & 0.85 & 0.3 & 12.9 & 
\ref{v392_per_only_v_x_big_disk_6100k_logscale}b,orange   \\
1.37 & 6100 & $1.7\times 10^{-7}$ & 0.85 & 0.3 & 14.7 & 
\ref{v392_per_only_v_x_big_disk_6100k_logscale}b,magenta  \\
1.37 & 6100 & $1.7\times 10^{-7}$ & 0.85 & 0.01 & 14.8 &  \\ 
1.37 & 6100 & $1\times 10^{-7}$ & 0.85 & 0.3 & 14.9 & 
\ref{v392_per_only_v_x_big_disk_6100k_logscale}b,cyan  \\
1.37 & 6100 & $1\times 10^{-7}$ & 0.85 & 0.01 & 14.9 &  \\
1.37 & 6100 & $1\times 10^{-8}$ & 0.85 & 0.01 & 15.1 & 
\ref{v392_per_only_v_x_big_disk_6100k_logscale}b,red  \\
1.37 & 6100 & $1\times 10^{-9}$ & 0.85 & 0.01 & 15.2 & 
\ref{v392_per_only_v_x_big_disk_6100k_logscale}b,gray  \\
\hline
1.37 & 4500 & $4.6\times 10^{-7}$ & 0.85 & 0.3 & 12.9 & \\
1.37 & 4500 & $1.7\times 10^{-7}$ & 0.85 & 0.3 & 15.3 & 
\ref{v392_per_only_v_x_big_disk_3500k_logscale}a,magenta  \\
1.37 & 4500 & $1.7\times 10^{-7}$ & 0.85 & 0.01 & 15.4 &  \\
1.37 & 4500 & $1\times 10^{-7}$ & 0.85 & 0.3 & 15.4 & 
\ref{v392_per_only_v_x_big_disk_3500k_logscale}a,cyan  \\
1.37 & 4500 & $1\times 10^{-7}$ & 0.85 & 0.01 & 15.5 &  \\
1.37 & 4500 & $1\times 10^{-8}$ & 0.85 & 0.01 & 15.9 & 
\ref{v392_per_only_v_x_big_disk_3500k_logscale}a,red  \\
1.37 & 4500 & $1\times 10^{-9}$ & 0.85 & 0.01 & 16.1 & 
\ref{v392_per_only_v_x_big_disk_3500k_logscale}a,gray  \\
\hline
1.37 & 3500 & $4.6\times 10^{-7}$ & 0.85 & 0.3 & 12.9 & \\
1.37 & 3500 & $1.7\times 10^{-7}$ & 0.85 & 0.3 & 15.6 & 
\ref{v392_per_only_v_x_big_disk_3500k_logscale}b,magenta  \\
1.37 & 3500 & $1.7\times 10^{-7}$ & 0.85 & 0.01 & 15.7 &  \\
1.37 & 3500 & $1\times 10^{-7}$ & 0.85 & 0.3 & 15.7 & 
\ref{v392_per_only_v_x_big_disk_3500k_logscale}b,cyan  \\
1.37 & 3500 & $1\times 10^{-7}$ & 0.85 & 0.01 & 16.0 &  \\
1.37 & 3500 & $1\times 10^{-8}$ & 0.85 & 0.01 & 16.9 & 
\ref{v392_per_only_v_x_big_disk_3500k_logscale}b,red  \\
1.37 & 3500 & $1\times 10^{-9}$ & 0.85 & 0.01 & 17.2 & 
\ref{v392_per_only_v_x_big_disk_3500k_logscale}b,gray 
\enddata
\tablenotetext{a}{
The $V$ magnitude at the right end of each model light curve
for the distance modulus in the $V$ band of $(m-M)_V=14.6$.
The inclination angle is assumed to be $i=20\arcdeg$
for a binary of $M_{\rm WD}=1.37 ~M_\sun$ (Ne3) and $M_2=1.0 ~M_\sun$.
}
\end{deluxetable}

\begin{figure*}
\epsscale{1.0}
\plottwo{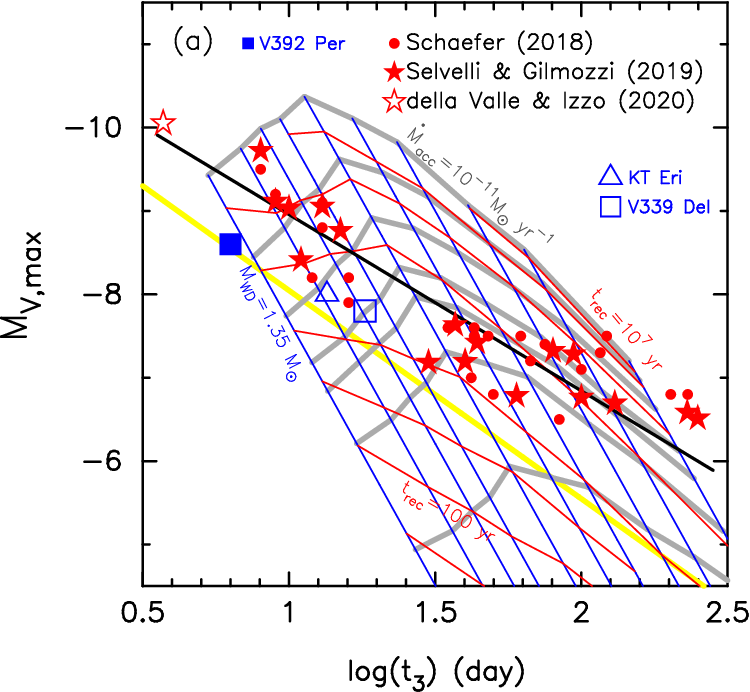}{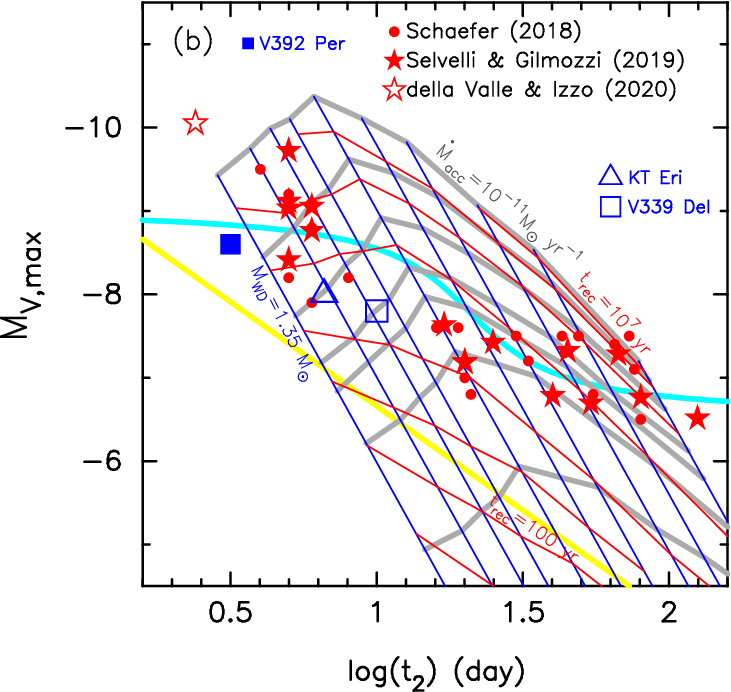}
\caption{
Theoretical maximum $V$ magnitude versus rate of decline (MMRD) diagram,
(a) $\log (t_3)$-$M_{V,\rm max}$ and (b) $\log (t_2)$-$M_{V,\rm max}$.
The blue lines indicate model equi-WD mass lines, from left to right,
1.35, 1.3, 1.25, 1.2, 1.1, 1.0, 0.9, 0.8, 0.7, and $0.6~M_\sun$;   
the thick solid gray lines denote model equi-mass accretion rate 
($\dot M_{\rm acc}$) lines, from lower to upper, $3\times 10^{-8}$,
$1\times 10^{-8}$, $5\times 10^{-9}$, $3\times 10^{-9}$, $1\times 10^{-9}$, 
$1\times 10^{-10}$, and $1\times 10^{-11} M_\sun$~yr$^{-1}$;
the red lines represent model equi-recurrence time lines, from lower to upper,
$t_{\rm rec}= 30$, 100, 300, 1000, 10000, $10^5$, $10^6$, and $10^7$~yr.
These lines are taken from \citet{hac20skhs} based on the optically thick
nova wind model \citep{kat94h}
and nuclear runaway model calculation of mass accretion onto each WD. 
The thick yellow line corresponds to the $x_0=2$ line, below which
the models are not valid \citep[see ][for details]{hac20skhs}.
We overplot V392 Per (filled blue square), 
filled red circles taken from ``Golden sample'' of \citet{schaefer18},
filled stars taken from \citet{sel19}, and open star (V1500~Cyg) 
taken from \citet{del20i}.
We further add two novae, KT Eri \citep[open blue triangle,][]{hac25kw}
and V339 Del \citep[open blue square, ][]{hac24km}. 
In panel (a), the thick solid black line indicates the empirical
line for the MMRD relation obtained by \citet{sel19}.
In panel (b), the thick solid cyan line represent the empirical MMRD
line obtained by \citet{del20i}.
\label{vmax_t3_vmax_t2_selvelli2019_schaefer2018_2fig}}
\end{figure*}

\subsection{WD mass, mass accretion rate, and recurrence period from
MMRD diagram}
\label{mmrd_diagram}

\citet{hac20skhs} presented theoretical maximum magnitude 
versus rate of decline (MMRD) diagrams based on their database 
of theoretical nova light curves, which are calculated 
using the optically thick wind model \citep{kat94h}. 
Figure \ref{vmax_t3_vmax_t2_selvelli2019_schaefer2018_2fig} shows 
their theoretical MMRD diagrams. 
These plots can be used to estimate the WD mass and 
mass accretion rate from the peak $V$ magnitude ($M_{V, \rm max}$) 
and the rate of decline ($t_3$ or $t_2$) of a nova. 
For example, 
the detailed light curve analysis for KT Eri \citep{hac25kw}
and V339 Del \citep{hac24km}
confirmed that this MMRD diagram gives very consistent binary 
parameters with those obtained from their light curve analysis. 

Figure \ref{vmax_t3_vmax_t2_selvelli2019_schaefer2018_2fig} also shows 
various novae that distribute in the middle of the diagram 
with a relatively large scatter (data are taken from 
\citet{schaefer18}, \citet{sel19}, and \citet{del20i}, with the distances
estimated based on the Gaia DR2). 
This large scatter stems from the difference in the binary parameters 
such as the WD mass and mass accretion rate. 

There are two well known empirical MMRD relations:  
The left panel shows the relation proposed by \citet{sel19} 
(thick solid black line) and the right panel does by \citet{del20i} 
(thick solid cyan line). 
These two relations are located in the middle of the distribution,
thus, it has been used to estimate the maximum magnitude
from each decline rate \citep[e.g.,][]{cho21sh, mur22dh}.  
The origin of this large scatter around the empirical MMRD relations
(black or cyan lines) 
%
has been argued \citep[e.g.,][]{schaefer18}, that is, 
whether or not these large scatters originate
essentially from the intrinsic nova properties. 
This question was answered by \citet{hac20skhs}; 
the scatter of each nova in the MMRD diagram is intrinsic as clearly shown in 
Figure \ref{vmax_t3_vmax_t2_selvelli2019_schaefer2018_2fig}, depending 
mainly on the WD mass and mass-accretion rate.

We plot the position of V392 Per in this figure (large filled blue square).  
The peak absolute $V$ is obtained to be $M_{V, \rm max}= 
m_{V, \rm max} - (m-M)_V = 6.0 - 14.6 = -8.6$ mag. 
The rate of decline, $t_2= 3$ or $t_3= 6.3$~days.
Comparing this position with theoretical equi-$M_{\rm WD}$,
equi-$\dot M_{\rm acc}$, and equi-$t_{\rm rec}$ lines, 
we obtain $M_{\rm WD}= 1.36~M_\sun$, 
$\dot M_{\rm acc}=6\times 10^{-11}~M_\sun$~yr$^{-1}$, and 
$t_{\rm rec}\sim 3\times 10^4$~yr from 
Figure \ref{vmax_t3_vmax_t2_selvelli2019_schaefer2018_2fig}a, but
$M_{\rm WD}= 1.37~M_\sun$, 
$\dot M_{\rm acc}=5\times 10^{-11}~M_\sun$~yr$^{-1}$, and 
$t_{\rm rec}\sim 4\times 10^4$~yr from 
Figure \ref{vmax_t3_vmax_t2_selvelli2019_schaefer2018_2fig}b. 

The positions of V392 Per in Figure
\ref{vmax_t3_vmax_t2_selvelli2019_schaefer2018_2fig} tell us that 
the WD mass is $M_{\rm WD}= 1.36$--$1.37 ~M_\sun$,
the mean mass accretion rate and recurrence time are 
$\dot M_{\rm acc}\sim (5$--$6)\times 10^{-11}~M_\sun$~yr$^{-1}$ and 
$t_{\rm rec}\sim (3$--$4)\times 10^4$~yr, respectively.
This range of the mass-accretion rate is consistent with the dwarf nova
nature of V392 Per before the 2018 nova outburst \citep[see, e.g.,][for
a review on dwarf novae]{osa96}.  We first adopt $M_{\rm WD}= 1.37 ~M_\sun$.
If this mass model does not satisfy the light curve, we will change
the WD mass to $M_{\rm WD}= 1.36 ~M_\sun$.

\subsection{Companion star}
\label{companion_star}

Based on the PanSTARS $grizY$ and 2MASS $JHK_{\rm s}$
spectral energy distribution (SED) fitting, 
\citet{mun20mm} obtained the companion mass to be
$M_2= 1.03 ~M_\sun$ (the effective temperature of $T_{\rm eff, 2}= 4740$ K),
$1.35 ~M_\sun$ ($4875$ K), and $1.92 ~M_\sun$ ($5915$ K), depending on
the reddening of $E(B-V)= 0.63$, $0.72$, and $1.18$,
respectively, from the Padova isochrones fitting \citep{bre12mg}
for the 2MASS plane \citep[color-magnitude diagram of
$(J - K_{\rm s})_0$-$M_{K_{\rm s}}$, see Figure 4 of ][]{mun20mm}.
Because we have already obtained/adopted the distance of $d= 3.45$ kpc and
the reddening $E(B-V)= 0.62$ in Section \ref{distance_reddening_orbit},
we adopt $M_2= 1.03 ~M_\sun$ and $T_{\rm eff, 2}= 4740$ K
(before the 2018 outburst) among the three. 
This companion mass is consistent with the fact that the mass-transfer 
should not be thermally unstable.  In other words, the mass ratio of
$M_2/M_{\rm WD} = 1.03/1.37 = 0.75 < 0.79$ does not result in a thermal
timescale mass transfer \citep[see, e.g., Equation (1) of][]{hac99kn},
where $0.79$ is the critical mass ratio for thermally unstable
mass transfer \citep[e.g.,][]{hac99kn}.
Therefore, we do not accept a $M_2=1.35 ~M_\sun$ subgiant because
it results in a thermally unstable mass transfer, as high as
$\dot{M}_{\rm acc} \gtrsim 1\times 10^{-7} ~M_\sun$ yr$^{-1}$,
for $M_{\rm WD}= 1.37 ~M_\sun$.  Such a
high mass transfer rate is not consistent with the dwarf nova nature
of V392 Per before the 2018 outburst.

For the effective temperature of the companion star, \citet{schaefer22a}
reported $T_{\rm eff, 2}= 6100\pm 330$ K ($M_2=1.04 ~M_\sun$, in quiescence) 
from the fluxes of Galex, APASS, Pan-STARRS, 2MASS, and WISE at different
epochs while \citet{mur22dh} proposed $T_{\rm eff, 2}= 5700\pm 400$ K
(before outburst) based on the spectra including the WISE mid-IR data.
In our modeling below, we adopt $M_2= 1.0 ~M_\sun$ and 
assume three cases of $T_{\rm eff,2}$, that is, 6100, 4500, and 3500 K, to
check whether or not our model light curve reproduces well the $V$ brightness,
just before and after the outburst.
Here, $T_{\rm eff, 2}= 6100$ K is the highest temperature satisfying both
Schaefer's and Murphy-Glaysher et al.'s estimates and 
$T_{\rm eff, 2}= 3500$ K is our trial case that reproduces the faintest
brightness $V\sim 17$ in quiescence before the 2018 outburst.
The $T_{\rm eff, 2}= 4500$ K is a middle between them and roughly close
to Munari et al.'s estimate of $T_{\rm eff, 2}= 4740$ K.

\subsection{Overall properties of the nova light curves}
\label{light_curve_summary}

A nova outburst starts from unstable hydrogen burning on a WD.
A hydrogen-rich envelope of the WD expands and emits strong winds
\citep[e.g.,][for a recent nova outburst calculation]{kat22sha}.
Free-free emission from the nova winds dominates the optical flux of a nova
\citep[e.g.,][]{gal76, enn77}.  \citet{hac06kb} modeled nova light curves
for free-free emission based on the optically thick winds calculated
by \citet{kat94h}, the $V$ flux of which can be simplified as
\begin{equation}
L_{V, \rm ff,wind} = A_{\rm ff} ~{{\dot M^2_{\rm wind}}
\over{v^2_{\rm ph} R_{\rm ph}}}.
\label{free-free_flux_v-band}
\end{equation}
This $V$ flux represents the flux of free-free emission from optically thin
plasma just outside the photosphere, and $\dot{M}_{\rm wind}$ is the
wind mass-loss rate, $v_{\rm ph}$ the velocity at the photosphere,
and $R_{\rm ph}$ the photospheric radius.  See \citet{hac20skhs} for
the derivation of this formula and the coefficient $A_{\rm ff}$.
In our $V$ light curve model, the total $V$ band flux is defined by 
the summation of the free-free (FF) emission luminosity
and the $V$ band flux of the photospheric luminosity $L_{\rm ph, WD}$
(assuming blackbody (BB)), i.e., FF+BB,
\begin{equation}
L_{V, \rm total} = L_{V, \rm ff,wind} + L_{V, \rm ph, WD}.
\label{luminosity_summation_flux_v-band}
\end{equation}
The photospheric $V$ band luminosity of the WD
is calculated from a blackbody with
$T_{\rm ph}$ and $L_{\rm ph}$ using a canonical response function
of the $V$ band filter, where $T_{\rm ph}$ and $L_{\rm ph}$ are
the photospheric temperature and luminosity, respectively.

Figure \ref{v392_per_only_v_x_big_disk_6100k_logscale}b plots the model
$V$ light curve (solid black line) of our $1.37 ~M_\sun$ WD with the
envelope chemical composition of neon nova 3 (Ne3: $X=0.65$, $Y=0.27$,
$Z=0.02$, $X_{\rm CNO}= 0.03$, and $X_{\rm Ne}=0.03$, where $X$ is the
hydrogen, $Y$ the helium, $Z$ the heavy elements, $X_{\rm CNO}$ the extra
carbon-nitrogen-oxygen, $X_{\rm Ne}$ the extra neon, all by mass weight).
We adopt the distance modulus in the $V$ band of $\mu_V \equiv (m-M)_V=14.6$.
Appendix \ref{optically_thick_wind_model} describes the details of
this light curve model.

Our model $V$ light curve follows well the $V$ observation except for
during day 7 to day 15.  The good agreement confirms that the
choice of a $1.37 ~M_\sun$ (Ne3) WD in Section \ref{mmrd_diagram} is
appropriate.  In other words, \citet{hac20skhs}'s theoretical MMRD diagram
gives a reasonable value of the WD mass if the maximum magnitude 
($M_{V, \rm max}$) and the decline rate ($t_2$ or $t_3$) are well
approximated by Equation (\ref{universal_decline_law}).

Our model light curve cannot explain the excess during day 7 to day 15,
which we attribute to a magnetic activity (See Section
\ref{secondary_maximum_plateau}).

\subsection{X-ray light curve}
\label{x-ray_light_curve}

Our model X-ray light curve is calculated from a blackbody with
$T_{\rm ph}$ and $L_{\rm ph}$ of our WD model using a 0.3-10 keV band filter.
We neglect absorption outside the photosphere.
Thus, the soft X-ray flux (thin colored lines in Figure 
\ref{v392_per_only_v_x_big_disk_6100k_logscale}b)
rapidly increases when the wind mass loss rate decreases to zero
on day 43 and starts to decay when steady hydrogen burning ends on day 65.

There are no X-ray data between the discovery date and day 83 as shown in
Figure \ref{v392_per_only_v_x_big_disk_6100k_logscale}a because of
the Sun constraint \citep{mur22dh}.
We suppose that the X-ray turn-off date, i.e., the end of the supersoft X-ray
source phase (SSS), is close to the first Swift X-ray observation after the
Sun constraint, i.e., day 84 \citep[see also the suggestion by][]{mur22dh}.
A sharp decrease of X-ray count rate on day 84--97 corresponds to the last
tail of the SSS phase \citep[e.g., ][]{mur22dh}.

Some our model X-ray light curves decay earlier than this date.
This difference can be explained if mass-accretion starts at a high rate
before/during the SSS phase (or starts even earlier than the SSS phase).
Recently, such a high mass-accretion rate is applied to the KT Eri model
to explain a long duration of the SSS phase \citep{hac25kw}.

Our nova evolution timescale is governed by 
a time-evolutionary sequence of the decreasing envelope mass of
\begin{equation}
{{d} \over {d t}} M_{\rm env} = \dot M_{\rm acc}
- \dot M_{\rm wind} - \dot M_{\rm nuc},
\label{nova_evoluion_eq}
\end{equation}
where $M_{\rm env}$ is the mass of a hydrogen-rich envelope on the WD, 
$\dot M_{\rm acc}$ the mass accretion rate onto the WD,
$\dot M_{\rm nuc}$ the mass decreasing rate of hydrogen-rich envelope
by nuclear (hydrogen) burning, and usually $\dot M_{\rm acc} \ll 
\dot M_{\rm wind}$, and $\dot M_{\rm acc} \ll \dot M_{\rm nuc}$ for
typical classical novae \citep[see, e.g.,][]{hac06kb}.
Therefore, the envelope mass is decreased by winds and nuclear burning.
A large amount of the envelope mass is lost mainly by winds because
of $\dot M_{\rm wind} \gg \dot M_{\rm nuc}$ in the early phase of
nova outbursts (see Equation (7) of \citet{hac06kb} for details).

Mass-accretion of a high rate prolongs the duration of the SSS phase,
because new fuel is supplied to hydrogen burning.
We are able to reproduce the end day (day 65) of the SSS phase by 
$\dot{M}_{\rm acc}= 1.7\times 10^{-7} ~M_\sun$ yr$^{-1}$ (thin magenta line
in Figure \ref{v392_per_only_v_x_big_disk_6100k_logscale}b). 
In this model, the wind phase and the X-ray turn-on time hardly change
because the wind mass-loss rate is much larger than the mass-accretion rate.

Figure \ref{v392_per_only_v_x_big_disk_6100k_logscale}b shows how
the X-ray light curve changes depending on the mass-accretion rate.
If we assume a smaller mass-accretion rate, the X-ray turnoff time
becomes somewhat earlier.  A larger mass-accretion rate can delay
the X-ray turnoff. 
For an extreme case of high mass accretion rate,
$\dot{M}_{\rm acc}\ge 4.6\times 10^{-7} ~M_\sun$ yr$^{-1}$,
hydrogen burns steadily and never stops, which is demonstrated by
the thin orange lines in Figure
\ref{v392_per_only_v_x_big_disk_6100k_logscale}b.

\subsection{Optical contributions of accretion disk and companion star}
\label{contribution_disk_companion}

\subsubsection{observational implications}
\label{observational_implications}

\citet{mun20mm}, \citet{schaefer22a},
and \citet{mur22dh} suggest a large contribution to the optical brightness
from the disk and companion star irradiated by the hydrogen-burning WD.
The post-outburst SEDs show hotter components compared with the pre-outburst
SEDs \citep[e.g.,][]{mun20mm}.
The $V$ brightness stopped the decline about 200 days after outburst and
remains stuck $\sim 2$ mag above the quiescent brightness before the 2018
outburst \citep{mun20mm}.  

\begin{figure}
\epsscale{1.15}
\plotone{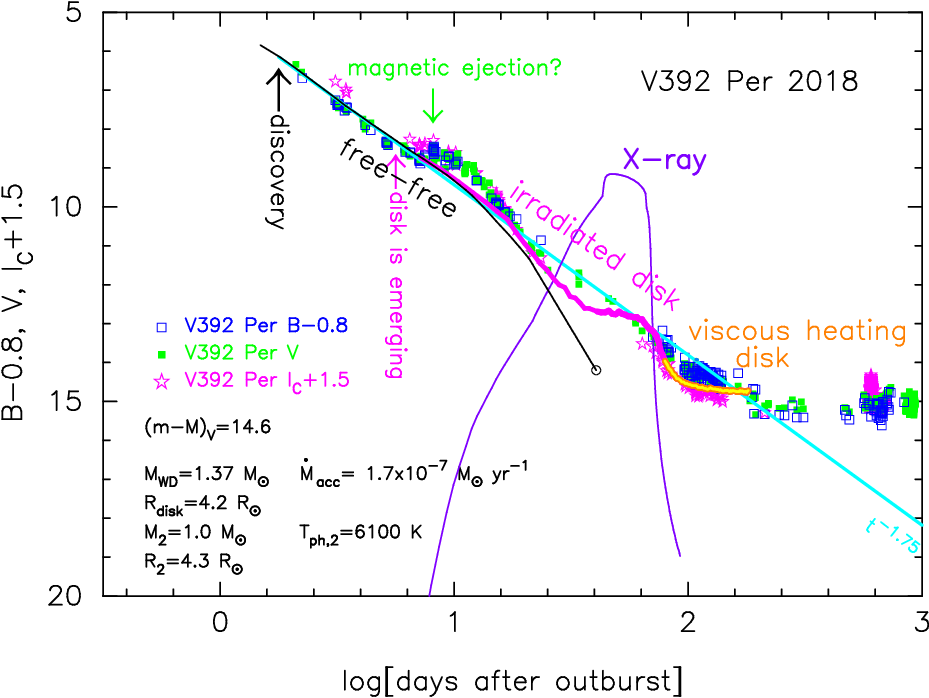}
\caption{
The $B$, $V$, and $I_{\rm C}$ light curves of V392 Per against a logarithmic
time.
We shift up $B$ by 0.8 mag but down $I_{\rm C}$ by 1.5 mag.
The $BVI_{\rm C}$ data are all taken from AAVSO.
The three light curves decay similarly.
The thin solid black line labeled ``free-free'' 
denotes our FF+BB model light curve of the $1.37 ~M_\sun$ WD (Ne3).
The magenta (labeled ``irradiated disk'') and orange+yellow (labeled 
``viscous heating disk'') lines denote our model $V$ light curve of
the magenta line in Figure
\ref{v392_per_only_v_x_big_disk_6100k_logscale}b, but the magenta part
is dominated by the irradiated disk and the orange+yellow part is
dominated by the viscous heating disk.
The purple line denotes the soft X-ray flux, which is the same as
the magenta line in Figure \ref{v392_per_only_v_x_big_disk_6100k_logscale}b.  
\label{bvi_light_curve}}
\end{figure}

We also point out that V392 Per shows a similarity in the declines of
$B$, $V$, and $I_{\rm C}$ light curves as shown
in Figures \ref{V392_per_kt_eri_v339_del_bvi_logscale}a
and \ref{bvi_light_curve}.
See also Figure 1 of \citet{mur22dh} or Figure 5 of \citet{cho21sh}.
This property is the same as in the classical nova KT Eri and
suggests that continuum emission dominates the optical and NIR
spectra of the novae all the time during the nova outburst
\citep[see Figure \ref{V392_per_kt_eri_v339_del_bvi_logscale}b of the
present paper and Figure 3a of ][for KT Eri]{hac25kw}, 
which is explained as the contribution of an irradiated (or 
viscous heating) bright disk except for during the early phase
near optical maximum (see Figure \ref{bvi_light_curve}). 
This is in contrast with typical classical novae, in which $B,V$ magnitudes
depart from the $I_{\rm C}$ magnitudes in the nebular phase 
\citep[see Figure \ref{V392_per_kt_eri_v339_del_bvi_logscale}c of
the present paper or Figure 11 of ][for V339 Del]{hac24km} owing to
contribution of [\ion{O}{3}] lines to the $B,V$ bands.

In KT Eri, a large irradiated disk contributes to the brightness 
in $B$, $V$, and $I_{\rm C}$ bands that makes a similar decline
in these bands \citep{hac25kw}. 
Such a large disk is plausible because
KT Eri has a long orbital period of 2.6 days and could host a large
accretion disk. Note that  
the recurrent nova U Sco, the orbital period of which is 1.23 days, 
is also observed to have a large disk in the recent 2022 outburst
\citep{mura24ki}.

\subsubsection{binary configuration}
\label{binary_configuration}

In what follows, we assume an irradiated disk even during the nova wind phase
and try to reproduce the $V$ light curve, the method of which is essentially
the same as that for KT Eri \citep{hac25kw}.  See 
\citet{hac01kb, hac03ka, hac03kb, hac03kc} for more details of
our irradiated disk models.


The photospheric emission from the bright disk dominates the spectra
of the nova.  The size of the disk is defined by
\begin{equation}
R_{\rm disk}= \alpha R_{\rm RL, 1},
\label{disk_radius_alpha}
\end{equation}
and the height of the disk at the edge is given by
\begin{equation}
H_{\rm disk}= \beta R_{\rm disk}.
\label{disk_height_beta}
\end{equation}
Here, $R_{\rm RL, 1}$ is the effective Roche-lobe
radius\footnote{The effective Roche lobe radius is
defined by the radius of a sphere of which the volume is the same as that
of the inner critical Roche lobe.  We adopt an approximate description 
proposed by \citet{egg83}.} for the WD component.

The size of the accretion disk is usually limited by the tidal limit,
broadly $\alpha \approx 0.85$ \citep[e.g., ][]{mura24ki},
although \citet{mura24ki} obtained
a large size of disk in the recurrent nova U Sco 2021 outburst,
the size of which corresponds to $\alpha = 1.3$ during the wind phase
(day 13 after optical maximum). 

The surface height $z$ of the disk at the equatorial distance 
$\varpi= \sqrt{x^2+y^2}$ from the center of the WD is assumed to be
\begin{equation}
z = \left({{\varpi} \over {R_{\rm disk}}}\right) H_{\rm disk},
\label{disk_shape_winds}
\end{equation}
during the wind phase, but
\begin{equation}
z = \left( {{\varpi} \over {R_{\rm disk}}}\right)^2 H_{\rm disk},
\label{disk_shape_sss}
\end{equation}
after the winds stop.

A flaring up disk was first studied in supersoft X-ray sources (SSSs)
in the Large Magellanic Cloud (LMC) by \citet{sch97mm}. 
They explained the orbital modulations of light curves  
with accretion disks whose edges are flaring up. 

\begin{figure}
\epsscale{1.15}
\plotone{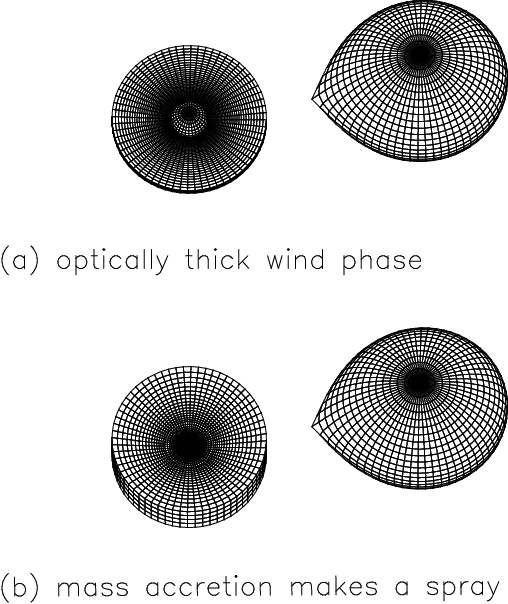}
\caption{
Geometric configuration models of our disk and companion star
in Figures \ref{v392_per_only_v_x_big_disk_6100k_logscale}b and
\ref{v392_per_only_v_x_big_disk_3500k_logscale}.
The masses of the WD and Roche-lobe-filling companion star
are $1.37 ~M_\sun$ and $1.0 ~M_\sun$, respectively.
The orbital period is $P_{\rm orb}= 3.23$ days.
The inclination angle of the binary is $i=20\arcdeg$.
The separation is $a= 12.26 ~R_\sun$ while their effective
Roche lobe radii are $R_{\rm RL, 1}= 4.98 ~R_\sun$ and
$R_{\rm RL, 2}= 4.31 ~R_\sun$.
The disk size is $R_{\rm disk}= 4.23 ~R_\sun$ ($=0.85~R_{\rm RL, 1}$).
We assume, in panel (a), the height of the disk edge to be 0.05 times
the disk size during the nova wind phase but, in panel (b), 
the edge height is 0.3 times the disk size after the winds stop.
The photospheric surfaces of the disk and companion star are irradiated
by the central hot WD. Such irradiation effects are all included in the
calculation of the $V$ light curve reproduction \citep[see][for the
partition of each surface and calculation method of irradiation]{hac01kb}.  
In panel (b), L1 stream from the companion impacts the disk edge and
makes a spray, which elevates the disk edge \citep{sch97mm}.
We also include the effect of viscous heating in the accretion disk
for a given mass-accretion rate \citep{hac01kb}.
\label{kt_eri_config}}
\end{figure}

In the present paper, we adopt $\alpha= 0.85$, which is close to the tidal
limit of an accretion disk \citep[see, e.g., ][]{mura24ki}. 
We also assume that the disk edge  height is $\beta= 0.05$ times
the disk size in the nova wind phase in Equation (\ref{disk_height_beta}),
because the surface of the disk is blown in the wind as suggested by
\citet{hac25kw}.  On the other hand, we assume $\beta\le 0.3$
after the winds stop.  
Such model configurations are shown in Figure \ref{kt_eri_config}.
For low mass-transfer rate such as
$\dot{M}_{\rm acc} \ll 1 \times 10^{-7} ~M_\sun$ yr$^{-1}$,
we adopt a small $\beta= 0.01$ 
because the mass-transfer rate is much smaller than those in the SSSs 
and the streaming impact could be too small to increase the edge height.


\subsubsection{model $V$ light curve}
\label{model_v_light_curve}

The $V$ brightness can be written by the summation of the free-free emission
luminosity and $V$ band fluxes of the photospheric luminosities, i.e.,
\begin{eqnarray}
L_{V, \rm total} &=& L_{V, \rm ff,wind} + L_{V, \rm ph, WD} \cr
  & & + L_{V, \rm ph, disk} + L_{V, \rm ph, comp},
\label{luminosity_summation_wd_disk_comp_v-band}
\end{eqnarray}
where $L_{V, \rm ph, disk}$ is the $V$ band flux from the disk,
and $L_{V, \rm ph, comp}$ the $V$ flux from the companion star. 
The irradiation effect is the main optical source in the disk
and companion star, but we also include the viscous heating of the disk
\citep{hac01kb}.  After hydrogen burning ends, the viscous heating
of the disk becomes important for a high mass-transfer rate system.

We assumed the effective temperature of the companion star to be
$T_{\rm eff, 2}= 6100$ K from estimates by \citet{schaefer22a}
and \citet{mur22dh} as introduced in Section \ref{companion_star}. 

For the $1.0 ~M_\sun$ companion star with the effective temperature
of $T_{\rm eff, 2}= 6100$ K,
we are able to broadly reproduce the $V$ light curve of V392 Per
by our $1.37 ~M_\sun$ WD model with the mass accretion rate of 
$\dot{M}_{\rm acc}= 1.7\times 10^{-7} ~M_\sun$ yr$^{-1}$ (magenta lines),
as plotted in Figure \ref{v392_per_only_v_x_big_disk_6100k_logscale}b.
At the end of the magenta line (on day 180), we start to reduce the mass
transfer rate from $-\dot{M}_2= 1.7\times 10^{-7}$
(magenta line) to $1.0\times 10^{-9} ~M_\sun$ yr$^{-1}$ (gray line).
The yellow arrows in Figure \ref{v392_per_only_v_x_big_disk_6100k_logscale}b
indicate the path in which we reduce $-\dot{M}_2$ from $1.7\times 10^{-7}$
to $1.0\times 10^{-9} ~M_\sun$ yr$^{-1}$.
To reproduce the $V$ brightness in quiescence (after day 210),
we may adopt $-\dot{M}_2 \lesssim 1\times 10^{-8} ~M_\sun$ yr$^{-1}$
for the mass-accretion rate to the WD through the disk. 
Here, we assume that the total mass of the binary is conserved, that is, 
$\dot{M}_{\rm acc} +\dot{M}_2 = 0$.

Our model $V$ light curve (thick magenta line) is $\Delta V\sim 0.5$ mag
fainter than the observation on day $\sim 34$ and day $\sim 47$,
although these are only two epoch $V$ observations between day 25 and 60.
Note that this period corresponds to the phase that the wind mass loss
rapidly weakens and the accretion disk changes its shape.  Thus, these
deviations can be explained by fluctuations during the transition 
from the wind phase to the SSS phase.  A similar behavior is also observed
in KT Eri as in Figures \ref{V392_per_kt_eri_v339_del_bvi_logscale}b
\citep[see also Figure 1 of ][]{hac25kw}.
Such fluctuations could originate from the time-variation 
of the disk shape (both in the $\alpha$ and $\beta$ parameters)  
during the transition, which is not included in our model.

\subsubsection{short summary of light curve fitting}
\label{short_summary_of_fitting}

Figure \ref{bvi_light_curve} summarizes main features of our model light curve.
In the early phase ($t \lesssim 7$ days), the $V$ light curve is dominated
by free-free emission from the nova wind (black line).
Then, the disk gradually emerges from the photosphere of the WD and
the irradiation effect of the disk becomes prominent after day $\sim 10$
(magenta line).

We did not model the secondary maximum, which we attribute to
a phenomenon driven by magnetic energy release
(see Section \ref{secondary_maximum_plateau}).

The nova wind stopped on day 43 for $\dot{M}_{\rm acc}= 1.7\times 10^{-7}
~M_\sun$ yr$^{-1}$.  An SSS phase could start from day $\sim 40$ and
ended on day 65.  This end day is consistent with the first detection
day (day 84) of soft X-rays,
that suggests a tail of dropping count rate (red crosses)
in Figure \ref{v392_per_only_v_x_big_disk_6100k_logscale}b.

The later phase (65 $\lesssim t \lesssim 200$ days) $V$ light curve is well
explained with the contribution of the viscous-heating accretion-disk.
This part corresponds to the orange+yellow line
in Figure \ref{bvi_light_curve}.  The mass accretion rate of
$\dot{M}_{\rm acc}\sim 1.7\times 10^{-7} ~M_\sun$ yr$^{-1}$ can satisfy 
both the $V$ brightness of $V\sim 14.7$ and the duration of the SSS phase
(until day 65). 

After day $\sim 200$, the $V$ brightness slightly declines to, 
and stays at, $V\sim 15.2$.
We reproduce this slight decline, if the $T_{2,\rm eff}= 6100$ K
remains the same but the mass accretion rate decreases from 
$\dot{M}_{\rm acc}\sim 1.7\times 10^{-7} ~M_\sun$ yr$^{-1}$
to $\dot{M}_{\rm acc}\sim 1\times 10^{-9} ~M_\sun$ yr$^{-1}$ or less.
We do not think that the companion temperature changes with time,
at least, during our light curve fitting with the 2018 outburst of
V392 Per.  This could be supported by the V brightness of $V=15.2$ just
before the 2018 outburst \citep[see Figure 2 of ][]{mur22dh}, because
we suppose that V392 Per comes back to the pre-outburst brightness,
$V=15.2$.

\subsection{disk and companion star post outburst}
\label{status_post_outburst}

\citet{mun20mm} extensively discussed that the post-nova level-off luminosity
\citep[e.g., $V=15.2\pm 0.1$ in Figure 2 of ][]{mur22dh}
is about 2 mag brighter than the pre-nova luminosity \citep[e.g., $V\sim 17$
in quiescence, Figure 2 of ][]{mur22dh}.  Munari et al. proposed an
idea that this brightness (sustained post-outburst brightness) is caused
by irradiation of the disk and companion star by the WD still burning at
the surface.  However, this idea cannot be supported by our steady hydrogen
burning model (orange line) in Figure  
\ref{v392_per_only_v_x_big_disk_6100k_logscale}b, because its brightness
keeps at $V\sim 13$ and does not decline to $V\sim 15$.

\citet{mur22dh} suggested a high mass-transfer rate post-nova to explain
this 2 mag brighter luminosity.  This is supported by our results
in Figures \ref{v392_per_only_v_x_big_disk_6100k_logscale}b and
\ref{v392_per_only_v_x_big_disk_3500k_logscale}a, if $T_{\rm eff,2}$ 
cools down from 6100 to 4500 K around on day $\sim 200$
with $\dot{M}_{\rm acc}=1.7\times 10^{-7} ~M_\sun$ yr$^{-1}$
being kept constant.  Then, the brightness changes from $V=14.7$ to $V=15.3$
around on day 200, as tabulated in Table \ref{post-outburst_parameters}.
After day 200, the brightness remains at $V=15.3$ if both the
$\dot{M}_{\rm acc}$ and $T_{\rm eff,2}$ remains the same (4500 K and
$1.7\times 10^{-7} ~M_\sun$ yr$^{-1}$).

However, the model $V$ light curve will not decay to $V\sim 17$, the
brightness in quiescence, even if
we reduce the mass transfer rate down to $\dot{M}_{\rm acc}=1\times 10^{-9}
~M_\sun$ yr$^{-1}$ for $T_{\rm eff,2}=6100$ K 
($V=15.2$, light gray line in Figure
\ref{v392_per_only_v_x_big_disk_6100k_logscale}b) or
$T_{\rm eff,2}=4500$ K 
($V=16.1$, light gray lines in Figure
\ref{v392_per_only_v_x_big_disk_3500k_logscale}a).  To reproduce
$V\sim 17$ in the pre-nova phase, we have to decrease the effective
temperature of the companion down to $T_{\rm eff,2}= 3500$ K
(Figure \ref{v392_per_only_v_x_big_disk_3500k_logscale}b).
This suggests that the effective temperature of the companion increased
from $\sim 3500$ K to $6100$ K before the outburst
\citep[see Figure 2 of ][]{mur22dh}.
This increase cannot be explained only by the irradiation effect
because hydrogen burning ended on day 65 (or, at least, day $\sim 80$)
much before the end of observation (day 1000 or later). 

Note that V392 Per is known as a dwarf nova variable with $V\sim 17-15$
\citep[Figure 2 of][]{mur22dh} and the pre-nova brightness had increased to
$V= 15.1$ about 200 days before the outburst.  
The post-outburst brightness $V=15.2\pm 0.1$ seems to be the same as
the pre-outburst brightness \citep[see Figure 2 of ][]{mur22dh}.
In this sense, the post-outburst $V$ brightness comes back to the 
pre-outburst brightness, not 2 mag brighter than the pre-outburst 
brightness.  However,
we must note that most of the pre-outburst photometry
was 1-2 magnitudes fainter than $V\sim 15$
and swings between $V=13.5$ and $V=17$.
The $V$ brightness could not be stable at $V=15.2$ all the day 
during 200 days before the 2018 outburst because there are only two epochs
of observation during this $\sim 200$ day period.


\begin{figure}
\gridline{\fig{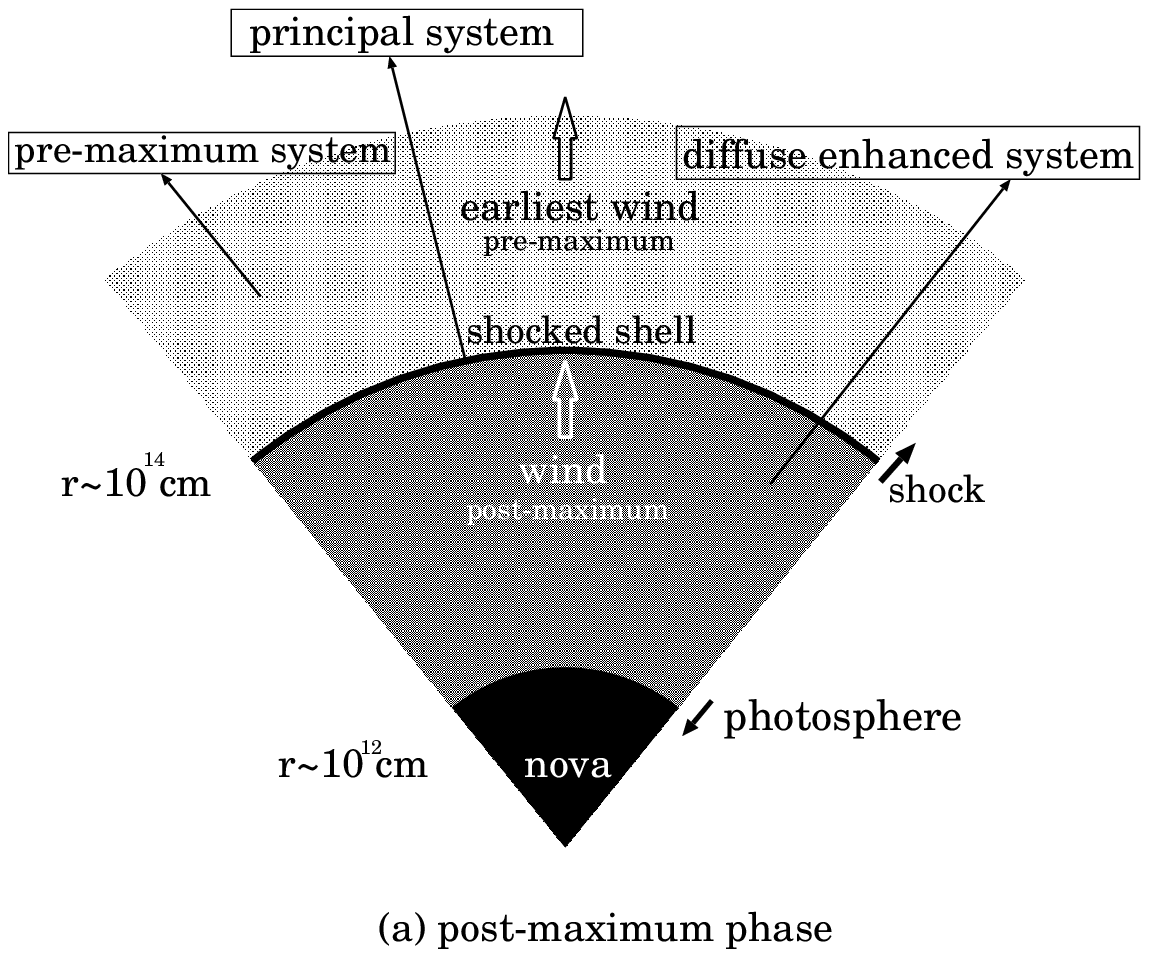}{0.45\textwidth}{}
          }
\gridline{\fig{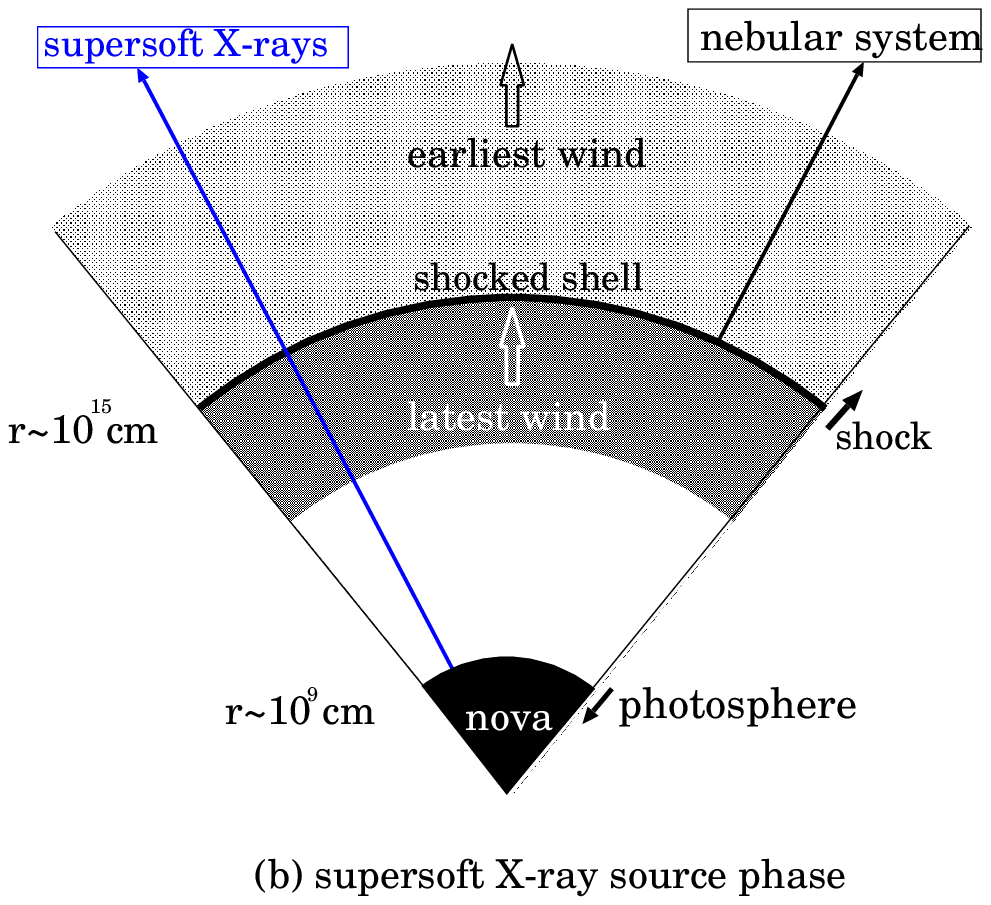}{0.45\textwidth}{}
          }
\caption{
Schematic illustration of a nova ejecta configuration of V392 Per
in the (a) post-maximum phase and (b) supersoft X-ray source (SSS) phase.
A shock wave arose just after the optical maximum and has already
moved far outside the WD photosphere (and the binary).
The shocked shell emits Balmer lines such as H$\alpha$.
This figure is taken from Figure 2(c) and (d) 
of \citet{hac23k} with a modification.
We assume that ejecta are spherically symmetric.
\label{v392_per_shock_configuration}}
\end{figure}

\section{Discussion}
\label{discussion}

\subsection{Gamma-ray emission}
\label{gamma_ray_flux}

As shown in Figure \ref{v392_per_only_v_x_big_disk_6100k_logscale}a,
the GeV gamma-ray flux is broadly correlated with the $V$ flux.
The nova GeV gamma-rays are considered to 
originate from a strong shock in the nova ejecta (internal shock)
or shock between ejecta and circumstellar matter (external shock) 
\citep[e.g.,][]{cho21ms}.  
Here, we first explain the formation and evolution of the internal shock,
then discuss the flux of gamma-rays from V392 Per.

\subsubsection{shock formation and its evolution}
\label{shock_evolution}

\citet{hac22k, hac23k} proposed a nova ejecta evolution model 
based on the fully self-consistent nova outburst model of \citet{kat22sha}. 
Hachisu \& Kato found that a strong shock naturally arises in nova ejecta
far outside the WD photosphere, and elucidated the origin of nova
absorption/emission line systems raised by \citet{mcl42}, as illustrated
in Figure \ref{v392_per_shock_configuration}.

The physical reason for a shock formation is as follows:
before the optical maximum, the photospheric wind velocity $v_{\rm ph}$
decreases toward maximum as the envelope expands. 
After the optical maximum, on the other hand, the photospheric wind velocity
turns to rapidly increase, so that the wind ejected later is catching up
the matter previously ejected.  Thus, matter will be compressed,
which causes a strong shock wave (reverse shock).  Such a trend of
the nova wind velocity evolution (decreasing and then increasing) has
also been confirmed with recent observation by \citet{ayd20ci}
for several classical novae.
The mass of the shocked shell ($M_{\rm shell}$) is increasing with time,
and reaches about $> 90$\% of the total ejecta mass.
Thus, a large part of nova ejecta is
eventually confined to the shocked shell \citep{hac22k}.

\citet{hac22k} interpreted that the principal absorption/emission line system
originates from the shocked shell.  On the other hand, the diffuse-enhanced
absorption/emission line system is from the inner wind, as illustrated 
in Figure \ref{v392_per_shock_configuration}a.
In V392 Per, we have P-Cygni profiles on day 2.1 \citep[Figure 9
of ][]{mur22dh}, which implies that $v_{\rm d}= v_{\rm wind}= 4600$
km s$^{-1}$ and $v_{\rm shock}= v_{\rm p}= 2500$ km s$^{-1}$.  Here,
$v_{\rm p}$ and $v_{\rm d}$ are the velocities of the principal and
diffuse-enhanced systems, respectively, and 
$v_{\rm wind}$ and $v_{\rm shock}$ are the velocities of the inner wind
and shock, respectively.  

Then, the temperature
just behind the shock is estimated to be
\begin{eqnarray}
kT_{\rm sh}& \sim & {3 \over 16} \mu m_p 
\left( v_{\rm wind} - v_{\rm shock} \right)^2 \cr
& \approx & 1.0 {\rm ~keV~} 
\left( {{v_{\rm wind} - v_{\rm shock}} \over  
{1000 {\rm ~km~s}^{-1}}} \right)^2,
\label{shock_kev_energy}
\end{eqnarray}
where $k$ is the Boltzmann constant,
$T_{\rm sh}$ is the temperature just after the shock
\citep[see, e.g.,][]{met14hv}, 
$\mu$ is the mean molecular weight ($\mu =0.5$ for hydrogen plasma),
and $m_p$ is the proton mass.
Substituting $v_{\rm shock}= v_{\rm p}=2500$ km s$^{-1}$ and 
$v_{\rm wind}= v_{\rm d}=4600$ km s$^{-1}$,
we obtain the post-shock temperature 
$k T_{\rm sh}\sim 4.4$ keV.

Mechanical energy of the wind is converted to thermal energy
by the reverse shock \citep{met14hv} as
\begin{eqnarray}
L_{\rm sh}& \sim & {{9}\over {32}} {\dot M}_{\rm wind} 
{{( v_{\rm wind} - v_{\rm shock} )^3} \over {v_{\rm wind}}} \cr
&=& 1.8\times 10^{37}{\rm ~erg~s}^{-1}
\left( {{{\dot M}_{\rm wind}} \over 
{10^{-4} ~M_\sun {\rm ~yr}^{-1}}} \right) \cr
 &  & \times
\left( {{{v_{\rm wind} - v_{\rm shock}} \over {1000{\rm ~km~s}^{-1}}}}
\right)^3
\left( {{{1000{\rm ~km~s}^{-1}} \over {v_{\rm wind}}} }\right). 
\label{shocked_energy_generation}
\end{eqnarray}
Substituting $\dot{M}_{\rm wind}= 2.0 \times 10^{-4} ~M_\sun$ yr$^{-1}$
from our $1.37 ~M_\sun$ WD model,
we obtain the post-shock energy of
$L_{\rm sh} \sim 7.2\times 10^{37}$ erg s$^{-1}$ a day after maximum (day 3).

The column density of hydrogen is estimated from
$M_{\rm shell}= 4 \pi R_{\rm sh}^2 \rho h_{\rm shell}$,
where $M_{\rm shell}$ is the shocked shell mass,
$\rho$ is the density
in the shocked shell, and $h_{\rm shell}$ the thickness of the shocked shell.
If we take an averaged velocity of shell $v_{\rm sh}=
v_{\rm shell}= v_{\rm shock}= 2500$ km s$^{-1}$,
the shock radius is calculated from $R_{\rm sh}(t)= v_{\rm shock}\times t$.
This reads
\begin{eqnarray}
N_{\rm H} & = & {{X \over m_p} {{ M_{\rm shell} }
\over {4 \pi R^2_{\rm sh}}}} \cr
 & \approx & 4.8\times 10^{22} {\rm ~cm}^{-2}
\left({X \over {0.5}}\right)
\left( {{M_{\rm shell}} \over {10^{-5} M_\sun}} \right)
\left( {{R_{\rm sh}} \over {10^{14} {\rm ~cm}}} \right)^{-2}
\cr
 & \approx & 6.4 \times 10^{20} {\rm ~cm}^{-2}
\left({X \over {0.5}}\right)
\left( {{M_{\rm shell}} \over {10^{-5} M_\sun}} \right) \cr
& & \times
\left( {{v_{\rm shell}} \over {1000 {\rm ~km~s}^{-1}}} \right)^{-2}
\left( {{t} \over {100~{\rm day}}} \right)^{-2}.
\label{column_density_hydrogen_time}
\end{eqnarray}
This gives $N_{\rm H}\approx 3\times 10^{22}$ cm$^{-2}$ for
$M_{\rm shell}= 1\times 10^{-6} ~M_\sun$,
$v_{\rm shell}=2500$ km s$^{-1}$, and $t=3$ days.
This $N_{\rm H}$ value is so large that hard X-rays from the shocked
shell could not be detected even if the Swift/XRT observed the nova,
although the Sun constraint prevented the observation.

\citet{hac23k} estimated the shock duration $\tau_{\rm shock}$ by
\begin{equation}
\tau_{\rm shock}= {{t_{\rm ws}} \over
{\left( 1- {{v_{\rm p}} \over {v_{\rm d}}}\right)}},
\label{duration_of_shock}
\end{equation}
where $t_{\rm ws}$ is the wind stopping time.
Substituting $v_{\rm sh}\approx v_{\rm p}=2500$ km s$^{-1}$
(principal system), $v_{\rm ph}\approx v_{\rm d}=4600$ km s$^{-1}$
(diffuse-enhanced system), and $t_{\rm ws}=41$ days (the wind duration
just after the shock arises) into Equation (\ref{duration_of_shock}),
we obtain the shock duration of $\tau_{\rm shock}= 41/0.4565= 90$ days.
Therefore, we expect hard X-ray emission until about $t_0+90$ days
in V392 Per, where $t_0$ is the day of
optical maximum.  Its duration is as long as the end of the SSS phase.

We also expect that the nebular emission line profiles such as [\ion{O}{3}] 
start the frozen-in on day $t_0+90$ days.
The [\ion{O}{3}] 4959+5007\AA\  nebular lines
started and completed the frozen-in during days 72--212 \citep{mur22dh}
while the [\ion{O}{3}] 4363\AA\  auroral line settled down to the frozen-in
state after day 82 \citep{mur22dh}.

\citet{mur22dh} analyzed the X-ray spectrum on day 83 ($=$day 84 in our
time) observed with
the Swift/XRT and decomposed it to be a combination of two components:
one is a blackbody component with the temperature of $k T = 62^{+17}_{-14}$
eV and the other is a collisionally excited thin thermal plasma component
with the temperature of $k T = 2.3^{+1.2}_{-0.5}$ keV.  The blackbody
component probably comes from the photosphere of the cooling WD just
after hydrogen burning ended.
If the latter component originates from the still alive shock in the ejecta,
its temperature is broadly consistent with our estimate of 
$k T_{\rm sh}\sim 4.4$ keV on day 3. 
They obtained another high temperature component of $> 5$ keV
after day $t_0+ 90$ days.  They suggested that V392 Per is
an intermediate polar and a $> 50$ keV component comes from accretion
flow shocks guiding by magnetic fields of $10^6 \lesssim B \lesssim 10^7$ G
on the WD after day $t_0+ 110$.

\subsubsection{gamma-ray flux}
\label{gamma-ray_flux}

\citet{alb22aa} obtained the mean flux of GeV gamma-ray (0.1--300 GeV) to be 
$L_{\gamma}= 5\times 10^{35}$ erg s$^{-1}$ for the distance of 
$d= 3.5$ kpc.
In our model, the shock energy generation is 
$L_{\rm sh}=7.2\times 10^{37}$ erg s$^{-1}$ on day 3.
The ratio of $L_{\gamma}/L_{\rm sh}= 5/720 \approx 0.01$,
about 1\% conversion rate, is consistent with
\begin{equation}
L_{\gamma}= \epsilon_{\rm nth} \epsilon_{\gamma} L_{\rm sh}
\lesssim 0.03 ~L_{\rm sh},
\end{equation}
where $\epsilon_{\rm nth}\lesssim 0.1$ is the fraction of the shocked
thermal energy to accelerate nonthermal particles, and
$\epsilon_{\gamma}\lesssim 0.1$ is the fraction
of this energy radiated in the Fermi/LAT band \citep[typically 
$\epsilon_{\rm nth} \epsilon_{\gamma} < 0.03$;][]{met15fv}.
Thus, our shock model reasonably explains the GeV gamma-ray fluxes
observed by the Fermi/LAT \citep{alb22aa}.

A broad correlation between $L_V$ and $L_{\gamma}$ in Figure 
\ref{v392_per_only_v_x_big_disk_6100k_logscale}a can be explained by
the luminosity dependence on the wind mass-loss rate, $L_V\propto 
(\dot{M}_{\rm wind})^2$ from Equation (\ref{free-free_flux_v-band}),
whereas the shock energy generation rate, 
$L_{\rm sh}\propto \dot{M}_{\rm wind}$ from Equation
(\ref{shocked_energy_generation}).
Thus, $L_{\rm sh}$ rapidly decreases as $\dot{M}_{\rm wind}$ decreases. 
The rise of gamma-ray flux may be closely related to the secondary maximum
(from day 7 to 15) that can be explained by a new ejection of wind 
(increase in the $\dot{M}_{\rm wind}$ and possibly in the $v_{\rm wind}$)
(see Section \ref{secondary_maximum_plateau} below).

It should be noted that, in our model, the shock luminosity is
as large as $L_{\rm sh}=7.2\times 10^{37}$ erg s$^{-1}$ on day 3 but does not
exceed the photospheric luminosity of $L_{\rm ph,BB} \sim 2\times 10^{38}$
erg s$^{-1}$ and the total optical luminosity of 
$L_{\rm total,FF+BB}\sim 7\times 10^{38}$ erg s$^{-1}$.
Therefore, the shock luminosity does not much contribute to the optical
light near/at maximum light.

\subsection{Secondary maximum: possible magnetic activity}
\label{secondary_maximum_plateau}

Our model light curve cannot explain the excess in the $V$ light curve during
day 7 to day 15 (Figure \ref{v392_per_only_v_x_big_disk_6100k_logscale}b).
The thick cyan line labeled $t^{-1.75}$  shows the universal decline law of
$L_V\propto t^{-1.75}$ and our FF+BB light curve follows well this line.
The FF+BB light curve (or universal decline law) is obtained by assuming
steady-state winds from the nova envelope. 
Because our model is based on the steady wind mass-loss,
this brightness excess suggests extra violent mass ejections.
It is interesting that the gamma-ray flux shows a similar jump (orange
diamonds in Figure \ref{v392_per_only_v_x_big_disk_6100k_logscale}a)
during day 7 to day 10.  This also indicates an additional mass ejection.
We regard this optical/gamma-ray enhancements as the secondary maxima as  
shown in V2491 Cyg, V1493 Aql, and V2362 Cyg.

\citet{hac09ka} extensively discussed the physical origin of
such secondary maxima, and suggested that strong magnetic fields on 
the WDs play a role for the violent mass ejection during the secondary maximum.
If the secondary maximum of V392 Per has the same origin as those novae,
its WD could have strong magnetic fields.   

Such a strong magnetic field in V392 Per is discussed by 
\citet{mur22dh}.  They examined the X-ray spectra and showed a 62 eV
blackbody component (tail of the SSS phase) on day 83
but a hard ($> 50$ keV) optically thin plasma component after day 100.
They attributed this hard component
to a shock in the accretion column on an intermediate-polar system 
having magnetic fields of $10^6 \le B \le 10^7$ G.

Thus, we may conclude that the enhancement around day 10
is attributed to a magnetic activity on the WD.

\section{Conclusions}
\label{conclusions}
Our $1.37 ~M_\sun$ WD (Ne3) model well reproduces the light curves of 
V392 Per (Figure \ref{bvi_light_curve}).
The main results are summarized as follows:
\begin{enumerate}
\item In the early phase ($t \lesssim 20$ days), our $V$ light curve consists
of free-free emission from the ejecta just outside the photosphere
plus blackbody emission from the WD photosphere (FF+BB).
\item Our model light curve cannot explain the small excess
(secondary maximum) during day 7 to day 15,
which we attribute to a violent magnetic activity \citep{hac09ka}.
This could be supported by \citet{mur22dh}'s suggestion that
V392 Per is an intermediate-polar with 
magnetic fields of $10^6 \lesssim B \lesssim 10^7$ G on the WD.
\item In the middle phase (20 days $\lesssim t \lesssim$ 65 days),
the $V$ light curve is dominated by the contribution
from an accretion disk irradiated by the hydrogen-burning WD.  
\item The supersoft X-ray light curve is calculated by blackbody flux 
from the hydrogen-burning WD.  The longer turnoff time of hydrogen burning
is reproduced if we assume a high mass accretion rate of 
$\dot{M}_{\rm acc}= 1.7\times 10^{-7} ~M_\sun$ yr$^{-1}$
from day 40 to day 200 after the outburst. 
\item In the later phase (65 $\lesssim t \lesssim 200$ days),
the $V$ light curve is dominated by the viscous heating accretion disk
with the mass accretion rate of 
$\dot{M}_{\rm acc}\sim 1.7\times 10^{-7} ~M_\sun$
yr$^{-1}$ together with the $\sim 1.0 ~M_\sun$ companion of
$T_{\rm eff,2}\sim 6100$ K.     
\item After day $\sim 200$, the $V$ brightness stays
at $V\sim 15.2\pm 0.1$, suggesting that the $V$ brightness comes back to
the pre-outburst brightness \citep[$V=15.1$, from $\sim 200$ days
before outburst, ][]{mur22dh}.
This can be reproduced either by decreasing $\dot{M}_{\rm acc}$ from
$\sim 1.7\times 10^{-7}$ to $\sim 1\times 10^{-9} ~M_\sun$ yr$^{-1}$ for
$T_{\rm eff,2}= 6100$ K or by decreasing $T_{\rm eff,2}$ from 6100 to 4500 K
for $\dot{M}_{\rm acc}= 1.7\times 10^{-7} ~M_\sun$ yr$^{-1}$.
\item In the quiescent phase before the outburst, V392 Per swings its $V$
brightness between $V\sim 17$ and 15 (sometimes up to 13.5).
To explain the faintest brightness of $V\sim 17$ in quiescence,
we must assume $T_{\rm eff,2}$ to be as low as 3500 K, much lower than 6100 K.
The origin of such variations in $T_{\rm eff,2}$ is unclear.
\item A nova ejecta is divided by the shock into three parts,
the outermost expanding gas (earliest wind before maximum),
shocked shell, and inner fast wind:  These three regions are responsible
for the pre-maximum, principal, and diffuse-enhanced absorption/emission
line systems \citep{mcl42}, respectively.
We interpret that the shock velocity $v_{\rm shock}$
corresponds to the velocity $v_{\rm p}$ of the principal system and
the inner wind velocity $v_{\rm wind}$ to the velocity $v_{\rm d}$ of the
diffuse-enhanced system.  The shock temperature is calculated to be
$k T_{\rm sh} \sim 4.4$ keV from equation (\ref{shock_kev_energy}),
assuming $v_{\rm p}= 2500$ km s$^{-1}$ and $v_{\rm d}= 4600$ km s$^{-1}$
from the observed spectra \citep{mur22dh}.
\item The shock energy generation rate is calculated to be
$L_{\rm sh}\sim 7.2\times 10^{37}$ erg s$^{-1}$
from Equation (\ref{shocked_energy_generation}).
The ratio of $L_{\gamma}/L_{\rm sh}\sim 0.01$ satisfies the theoretical
request \citep[$L_{\gamma}/L_{\rm sh} \lesssim 0.03$, ][]{met15fv}.
Here the observed GeV gamma-ray energy is $L_{\gamma}\sim 5\times 10^{35}$
erg s$^{-1}$ \citep{alb22aa}.  This supports our shock model as an
origin of the gamma-rays in V392 Per.
\item We obtain the distance modulus in the $V$ band  
to be $\mu_V\equiv (m-M)_V=14.6\pm0.2$, 
applying the time-stretching method to the $V$ light
curves of V392 Per, LV Vul, KT Eri, and V339~Del. 
The distance is $d=3.45\pm0.4$~kpc for the reddening
of $E(B-V)=0.62\pm0.02$ (Appendix \ref{time_stretching_method}). 
\item The theoretical maximum magnitude versus rate of decline diagram
\citep{hac20skhs} gives a consistent WD mass of $1.37 ~M_\sun$ (Ne3)
obtained from the $V$ light curve fitting.
The recurrence period and mass accretion rate are roughly estimated to be
$\sim 4\times 10^4$ yr and $\sim 5\times 10^{-11} ~M_\sun$ yr$^{-1}$,
respectively.
\end{enumerate}

\begin{acknowledgments}
     We thank
the American Association of Variable Star Observers
(AAVSO) and the Variable Star Observers League of Japan (VSOLJ)
for the archival data of V392 Per.
We are also grateful to the anonymous referee for useful comments
that improved the manuscript.
\end{acknowledgments}

\vspace{5mm}
\facilities{Swift(XRT), AAVSO, SMARTS, VSOLJ}


\appendix


\begin{figure*}
\gridline{\fig{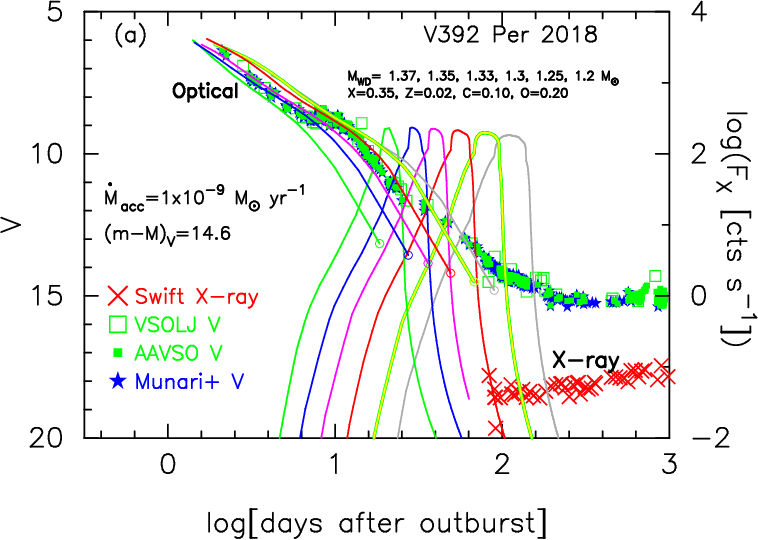}{0.45\textwidth}{}
          \fig{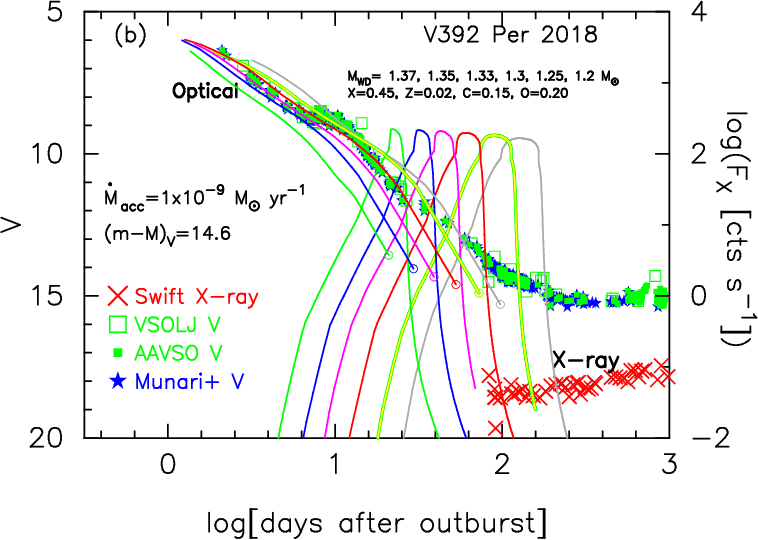}{0.45\textwidth}{}
          }
\gridline{
          \fig{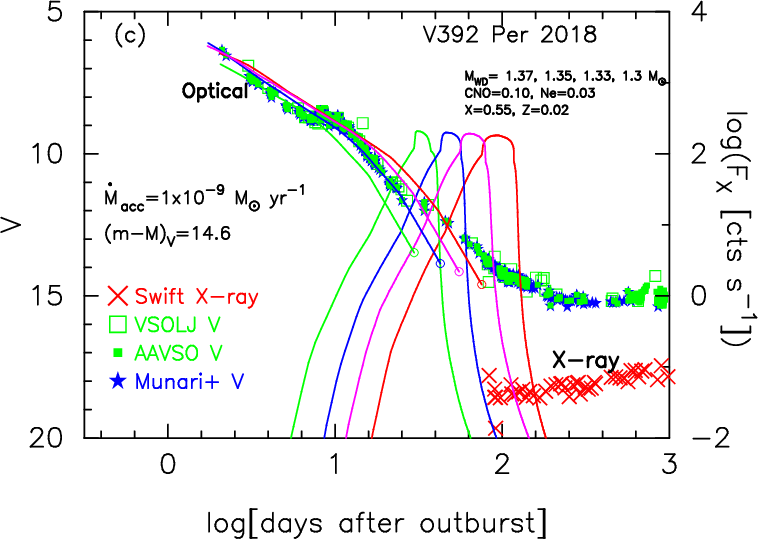}{0.45\textwidth}{}
          \fig{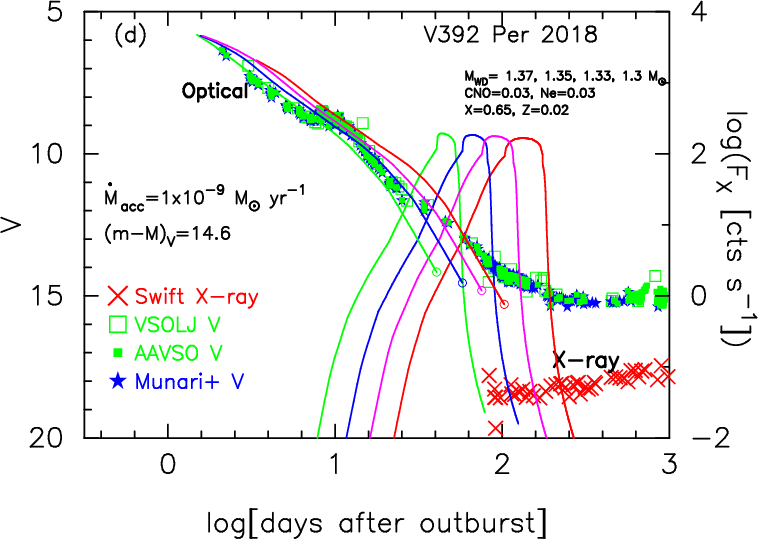}{0.45\textwidth}{}
          }

\caption{
Our FF+BB optical $V$ light curves for the
distance modulus in the $V$ band of $(m-M)_V=14.6$, and
our X-ray (0.3-10.0 keV) light curves, 
for different sets of WD mass and chemical composition.  
We add the same $V$ and X-ray count rate
data as those in Figure \ref{v392_per_only_v_x_big_disk_6100k_logscale}.
The mass-accretion rate onto the WDs
is fixed to be $\dot{M}_{\rm acc}= 1\times 10^{-9} M_\sun$ yr$^{-1}$
for all models.
(a) The chemical composition of the envelope is CO nova 2 (CO2):
$1.2~M_\sun$ (gray lines),  $1.25~M_\sun$ (yellow+green), 
$1.3~M_\sun$ (red), $1.33~M_\sun$ (magenta), 
$1.35~M_\sun$ (blue), and $1.37~M_\sun$ (green) WDs. 
(b) CO nova 3 (CO3): $1.2~M_\sun$ (gray), $1.25~M_\sun$ (yellow+green),
$1.3~M_\sun$ (red), $1.33~M_\sun$ (magenta), $1.35~M_\sun$ (blue),
and $1.37~M_\sun$ (green) WDs.
(c) Neon nova 2 (Ne2): $1.3~M_\sun$ (red), $1.33~M_\sun$ (magenta), 
$1.35~M_\sun$ (blue), and $1.37~M_\sun$ (green) WDs. 
(d) Neon nova 3 (Ne3): $1.3~M_\sun$ (red), $1.33~M_\sun$ (magenta), 
$1.35~M_\sun$ (blue), and $1.37~M_\sun$ (green) WDs.
\label{individual_v392_per_v_chemical_logscale_no2}}
\end{figure*}

\section{White Dwarf Models of Optical and Supersoft X-ray Light Curves}
\label{optically_thick_wind_model}

We present multiwavelength light curves based on the optically thick wind
model of novae, and constrain the range of possible white dwarf (WD) masses.
Here, we show the light curves only for free-free plus photospheric (FF+BB)
luminosity that does not include the contribution from a disk
and companion star.
Figure \ref{individual_v392_per_v_chemical_logscale_no2} shows our
FF+BB model $V$ light curves as well as supersoft X-ray light curves for
selected models. 

\citet{hac06kb, hac15k}
calculated many free-free emission light curves for novae with various WD
masses and chemical compositions based on \citet{kat94h}'s nova wind model. 
We call such a light curve model ``FF+BB'' (see Equations
(\ref{free-free_flux_v-band}) and 
(\ref{luminosity_summation_flux_v-band})).
The absolute magnitude of each FF+BB model light curve
has been calibrated with several
novae with a known distance modulus in the $V$ band 
\citep{hac15k, hac20skhs}.  This can be done by fixing
the coefficient $A_{\rm ff}$ in Equation (\ref{free-free_flux_v-band}).
On the other hand,
our model X-ray flux (0.3-10.0 keV) is calculated from the model WD
photosphere with blackbody assumption for the photospheric temperature
$T_{\rm ph}$ and photospheric radius $R_{\rm ph}$ \citep{kat94h}.  
These two (FF+BB and X-ray) model light curves have reproduced
the decay trends of various nova light curves.

\subsection{CO novae 2 (CO2)}
\label{co_novae_2}
Figure \ref{individual_v392_per_v_chemical_logscale_no2}a shows our model
light curves of $1.2~M_\sun$ (gray lines), $1.25~M_\sun$ (yellow+green),
$1.3~M_\sun$ (red), $1.33~M_\sun$ (magenta),
$1.35~M_\sun$ (blue), and $1.37~M_\sun$ (green) WDs
for the chemical composition of CO nova 2 \citep[CO2;][]{hac06kb},
i.e., $X=0.35$, $Y=0.33$, $Z=0.02$, $X_{\rm C}=0.10$, $X_{\rm O}=0.20$,
$X_{\rm Ne}=0.0$ by mass weight.  Here, $X_{\rm C}$, $X_{\rm O}$,
and $X_{\rm Ne}$ are the extra carbon, oxygen, and neon.
These extra carbon, oxygen, and neon indicate the degree of mixing
between the WD core and hydrogen-rich envelope \citep{hac06kb}.  
The corresponding FF+BB light curves are numerically tabulated in Table
\ref{light_curves_of_novae_co2} for 1.25, 1.3, 1.33, 1.35, and 1.37
$~M_\sun$ WDs.  The supersoft X-ray fluxes are not tabulated.
The other WD mass ($M_{\rm WD} \le 1.2 ~M_\sun$) cases were already
tabulated in \citet{hac10k}.

We select a best-fit model of the $1.33 ~M_\sun$ (magenta lines) WD
among the six WD mass models both for the $V$ and X-ray light curves.
To fit our X-ray flux with the soft X-ray decline on day 84, however,
we require a high mass-accretion rate on to the WD such as
$\dot{M}_{\rm acc}\sim 1.7\times 10^{-7} ~M_\sun$ yr$^{-1}$ and
extend the duration of the SSS phase (hydrogen burning), as shown
in Figure \ref{v392_per_only_v_x_big_disk_6100k_logscale}b.

\subsection{CO novae 3 (CO3)}
\label{co_novae_3}
Figure \ref{individual_v392_per_v_chemical_logscale_no2}b depicts
the light curves of
$1.2~M_\sun$ (gray lines), $1.25~M_\sun$ (yellow+green),
$1.3~M_\sun$ (red), $1.33~M_\sun$ (magenta),
$1.35~M_\sun$ (blue), and $1.37~M_\sun$ (green) WDs
for the chemical composition of CO nova 3 (CO3),
i.e., $X=0.45$, $Y=0.18$, $Z=0.02$, $X_{\rm C}=0.15$, $X_{\rm O}=0.20$,
$X_{\rm Ne}=0.0$.
The corresponding FF+BB $V$ light curves are numerically tabulated in Table
\ref{light_curves_of_novae_co3} for 1.25, 1.3, 1.33, 1.35, and 1.37
$~M_\sun$ WDs.  
The other WD mass ($M_{\rm WD} \le 1.2 ~M_\sun$) cases were 
tabulated in \citet{hac16k}.
We select a best-fit model of the $1.33 ~M_\sun$ (magenta lines) WD
among the six WD mass models both for the $V$ and X-ray light curves.

\subsection{Neon novae 2 (Ne2)}
\label{ne_novae_2}
Figure \ref{individual_v392_per_v_chemical_logscale_no2}c shows
$1.3~M_\sun$ (red lines), $1.33~M_\sun$ (magenta),
$1.35~M_\sun$ (blue), and $1.37~M_\sun$ (green) WDs
for the chemical composition of Neon nova 2 (Ne2),
i.e., $X=0.55$, $Y=0.30$, $Z=0.02$, $X_{\rm C}=0.0$, $X_{\rm O}=0.10$,
$X_{\rm Ne}=0.03$.
The corresponding FF+BB $V$ light curves are numerically tabulated in Table
\ref{light_curves_of_novae_ne2} for 1.33, 1.35, and 1.37 $~M_\sun$ WDs.  
The other WD mass ($M_{\rm WD} \le 1.3 ~M_\sun$) cases were 
tabulated in \citet{hac10k}.
We select a best-fit model of the $1.35 ~M_\sun$ (blue lines) WD
among the four WD mass models both for the $V$ and X-ray light curves.

\subsection{Neon novae 3 (Ne3)}
\label{ne_novae_3}
Figure \ref{individual_v392_per_v_chemical_logscale_no2}d shows
$1.3~M_\sun$ (red lines), $1.33~M_\sun$ (magenta),
$1.35~M_\sun$ (blue), and $1.37~M_\sun$ (green) WDs
for the chemical composition of Neon nova 3 (Ne3),
i.e., $X=0.65$, $Y=0.27$, $Z=0.02$, $X_{\rm C}=0.0$, $X_{\rm O}=0.03$,
$X_{\rm Ne}=0.03$.
The corresponding FF+BB light curves are numerically tabulated in Table
\ref{light_curves_of_novae_ne3} for 1.33, 1.35, and 1.37 $~M_\sun$ WDs.  
The other WD mass ($M_{\rm WD} \le 1.3 ~M_\sun$) cases were 
tabulated in \citet{hac16k}.
We select a best-fit model of the $1.37 ~M_\sun$ (green lines) WD
among the four WD mass models both for the $V$ and X-ray light curves.

\citet{mun18o} classified V392 Per as a neon nova based on the
strong neon line [\ion{Ne}{5}] 3426\AA\  as well as 
[\ion{Ne}{3}] 3869\AA.  Therefore, we may conclude that V392 Per hosts
the WD of mass between $1.35$ and $1.37 ~M_\sun$ assuming a typical neon
nova composition (Ne2 or Ne3).  This WD mass range is consistent with
our MMRD diagram analysis in Section \ref{mmrd_diagram}.



\startlongtable
\begin{deluxetable}{llllll}
\tabletypesize{\scriptsize}
\tablecaption{Free-free plus photospheric (FF+BB) $V$ Light Curves of
CO Novae 2 (CO2)\tablenotemark{a}
\label{light_curves_of_novae_co2}}
\tablewidth{0pt}
\tablehead{
\colhead{$m_{\rm ff}$} &
\colhead{1.25$M_\sun$} &
\colhead{1.3$M_\sun$} &
\colhead{1.33$M_\sun$} &
\colhead{1.35$M_\sun$} &
\colhead{1.37$M_\sun$} \\
\colhead{(mag)} &
\colhead{(day)} &
\colhead{(day)} &
\colhead{(day)} &
\colhead{(day)} &
\colhead{(day)} 
}
\colnumbers
\startdata
  5.000     & 0.0 &  &  &      &      \\
  5.250     & 0.8787     & 0.9370     & 0.6069     &  &      \\
  5.500     &  1.900     &  1.594     &  1.441     & 0.8021     &  \\
  5.750     &  2.498     &  1.979     &  2.145     &  1.457     & 0.8480  \\
  6.000     &  2.984     &  2.347     &  2.465     &  1.816     &  1.217  \\
  6.250     &  3.455     &  2.730     &  2.795     &  2.123     &  1.560  \\
  6.500     &  3.942     &  3.115     &  3.129     &  2.407     &  1.816  \\
  6.750     &  4.436     &  3.530     &  3.447     &  2.676     &  2.076  \\
  7.000     &  4.970     &  3.934     &  3.783     &  2.961     &  2.314  \\
  7.250     &  5.526     &  4.321     &  4.104     &  3.248     &  2.547  \\
  7.500     &  6.093     &  4.738     &  4.392     &  3.480     &  2.772  \\
  7.750     &  6.654     &  5.175     &  4.705     &  3.727     &  2.977  \\
  8.000     &  7.245     &  5.619     &  5.030     &  3.985     &  3.149  \\
  8.250     &  7.874     &  6.094     &  5.337     &  4.230     &  3.322  \\
  8.500     &  8.560     &  6.610     &  5.673     &  4.471     &  3.493  \\
  8.750     &  9.373     &  7.205     &  6.059     &  4.739     &  3.669  \\
  9.000     &  10.36     &  7.986     &  6.503     &  5.040     &  3.875  \\
  9.250     &  11.55     &  8.914     &  7.052     &  5.405     &  4.112  \\
  9.500     &  13.01     &  10.14     &  7.700     &  5.836     &  4.377  \\
  9.750     &  14.86     &  11.53     &  8.543     &  6.388     &  4.726  \\
  10.00     &  17.01     &  13.14     &  9.521     &  7.055     &  5.123  \\
  10.25     &  19.40     &  14.84     &  10.67     &  7.850     &  5.613  \\
  10.50     &  21.73     &  16.48     &  11.90     &  8.731     &  6.152  \\
  10.75     &  24.11     &  18.15     &  13.03     &  9.612     &  6.725  \\
  11.00     &  26.65     &  19.85     &  14.22     &  10.48     &  7.343  \\
  11.25     &  29.08     &  21.47     &  15.49     &  11.41     &  7.988  \\
  11.50     &  31.21     &  22.77     &  16.68     &  12.34     &  8.618  \\
  11.75     &  33.10     &  24.14     &  17.87     &  13.22     &  9.228  \\
  12.00     &  35.10     &  25.59     &  19.09     &  14.12     &  9.833  \\
  12.25     &  37.22     &  27.13     &  20.32     &  15.05     &  10.49  \\
  12.50     &  39.47     &  28.76     &  21.58     &  16.05     &  11.17  \\
  12.75     &  41.85     &  30.48     &  22.91     &  17.06     &  11.86  \\
  13.00     &  44.37     &  32.31     &  24.33     &  18.40     &  12.60  \\
  13.25     &  47.04     &  34.24     &  25.82     &  19.83     &  13.39  \\
  13.50     &  49.86     &  36.29     &  27.41     &  21.06     &  14.23  \\
  13.75     &  52.86     &  38.46     &  29.09     &  22.35     &  15.12  \\
  14.00     &  56.03     &  40.76     &  30.87     &  23.73     &  16.06  \\
  14.25     &  59.39     &  43.20     &  32.75     &  25.19     &  17.06  \\
  14.50     &  62.95     &  45.78     &  34.75     &  26.73     &  18.12  \\
  14.75     &  66.72     &  48.51     &  36.86     &  28.36     &  19.24  \\
  15.00     &  70.72     &  51.40     &  39.10     &  30.09     &  20.42  \\
\hline
X-ray\tablenotemark{b}
& 40.1 & 22.0 & 12.9  & 9.8 & 7.1  \\
\hline
$\log f_{\rm s}$\tablenotemark{c}
 & $-0.52$ & $-0.67$ & $-0.82$  & -0.95 & $-1.10$  \\
\hline
$M_{\rm w}$\tablenotemark{d}
 & $+0.2$ & $-0.1$ & $-0.4$ & $-0.7$  & $-1.1$  \\
\enddata
\tablenotetext{a}{The chemical composition of the envelope is assumed
to be that of CO nova 2 in Table 2 of \citet{hac16k}.}
\tablenotetext{b}{Duration of supersoft X-ray phase in units of days 
for $\dot M_{\rm acc}=1\times 10^{-9} ~M_\sun$ yr$^{-1}$.}
\tablenotetext{c}{Stretching factor with respect to the LV Vul
observation in Figure 9 of \citet{hac25kw}.}
\tablenotetext{d}{Absolute magnitudes at the bottom point (open circles)
of FF+BB $V$ light curve in Figure
\ref{individual_v392_per_v_chemical_logscale_no2}a
by assuming $(m-M)_V = 14.6$  (V392 Per).  The absolute $V$ magnitude
is calculated from $M_V=m_{\rm ff} -15.0 + M_{\rm w}$.}
\end{deluxetable}



\startlongtable
\begin{deluxetable}{llllll}
\tabletypesize{\scriptsize}
\tablecaption{FF+BB $V$ Light Curves of CO Novae 3 (CO3)
\label{light_curves_of_novae_co3}}
\tablewidth{0pt}
\tablehead{
\colhead{$m_{\rm ff}$} &
\colhead{1.25$M_\sun$} &
\colhead{1.3$M_\sun$} &
\colhead{1.33$M_\sun$} &
\colhead{1.35$M_\sun$} &
\colhead{1.37$M_\sun$} \\
\colhead{(mag)} &
\colhead{(day)} &
\colhead{(day)} &
\colhead{(day)} &
\colhead{(day)} &
\colhead{(day)} 
}
\colnumbers
\startdata
  4.500     & 0.0 &  &  &      &      \\
  4.750     & 0.5980     & 0.6353     & 0.6086     &  &      \\
  5.000     &  1.395     &  1.374     &  1.349     & 0.6650     &      \\
  5.250     &  2.327     &  2.039     &  2.016     &  1.279     &      \\
  5.500     &  2.936     &  2.613     &  2.422     &  1.760     &  \\
  5.750     &  3.456     &  3.042     &  2.809     &  2.141     & 0.4987  \\
  6.000     &  3.981     &  3.484     &  3.181     &  2.518     & 0.9672  \\
  6.250     &  4.485     &  3.897     &  3.521     &  2.846     &  1.351  \\
  6.500     &  5.009     &  4.306     &  3.843     &  3.141     &  1.629  \\
  6.750     &  5.540     &  4.738     &  4.188     &  3.445     &  1.909  \\
  7.000     &  6.103     &  5.126     &  4.559     &  3.714     &  2.177  \\
  7.250     &  6.657     &  5.550     &  4.859     &  3.966     &  2.376  \\
  7.500     &  7.256     &  6.000     &  5.164     &  4.230     &  2.582  \\
  7.750     &  7.864     &  6.427     &  5.468     &  4.498     &  2.806  \\
  8.000     &  8.519     &  6.883     &  5.800     &  4.734     &  2.983  \\
  8.250     &  9.231     &  7.378     &  6.170     &  4.994     &  3.158  \\
  8.500     &  10.00     &  7.937     &  6.594     &  5.276     &  3.348  \\
  8.750     &  11.03     &  8.664     &  7.129     &  5.603     &  3.553  \\
  9.000     &  12.21     &  9.518     &  7.749     &  5.993     &  3.792  \\
  9.250     &  13.70     &  10.58     &  8.527     &  6.459     &  4.069  \\
  9.500     &  15.40     &  11.83     &  9.430     &  7.042     &  4.406  \\
  9.750     &  17.49     &  13.37     &  10.50     &  7.731     &  4.807  \\
  10.00     &  19.89     &  15.06     &  11.70     &  8.561     &  5.291  \\
  10.25     &  22.55     &  16.67     &  12.99     &  9.439     &  5.851  \\
  10.50     &  24.90     &  18.37     &  14.20     &  10.32     &  6.442  \\
  10.75     &  27.39     &  20.20     &  15.48     &  11.26     &  7.045  \\
  11.00     &  29.56     &  21.82     &  16.60     &  12.25     &  7.685  \\
  11.25     &  31.40     &  23.16     &  17.60     &  13.15     &  8.359  \\
  11.50     &  33.34     &  24.59     &  18.67     &  14.08     &  8.988  \\
  11.75     &  35.40     &  26.09     &  19.80     &  15.05     &  9.631  \\
  12.00     &  37.58     &  27.69     &  20.99     &  16.01     &  10.31  \\
  12.25     &  39.88     &  29.38     &  22.26     &  16.98     &  11.00  \\
  12.50     &  42.33     &  31.17     &  23.60     &  18.02     &  11.71  \\
  12.75     &  44.92     &  33.07     &  25.02     &  19.12     &  12.61  \\
  13.00     &  47.66     &  35.08     &  26.53     &  20.28     &  13.61  \\
  13.25     &  50.57     &  37.21     &  28.12     &  21.51     &  14.64  \\
  13.50     &  53.65     &  39.46     &  29.81     &  22.81     &  15.93  \\
  13.75     &  56.91     &  41.85     &  31.60     &  24.19     &  16.89  \\
  14.00     &  60.37     &  44.38     &  33.49     &  25.65     &  17.92  \\
  14.25     &  64.02     &  47.06     &  35.50     &  27.20     &  19.01  \\
  14.50     &  67.90     &  49.90     &  37.63     &  28.84     &  20.17  \\
  14.75     &  72.01     &  52.91     &  39.88     &  30.57     &  21.39  \\
  15.00     &  76.36     &  56.10     &  42.26     &  32.41     &  22.68  \\
\hline
X-ray
& 51.5 & 29.1 & 18.9  & 11.5 & 6.9  \\
\hline
$\log f_{\rm s}$
 & $-0.49$ & $-0.64$ & $-0.79$  & $-0.92$ & $-1.07$  \\
\hline
$M_{\rm w}$
 & 0.3 & 0.0 & $-0.2$ & $-0.5$  & $-1.0$  \\
\enddata
\end{deluxetable}



\startlongtable
\begin{deluxetable}{llll}
\tabletypesize{\scriptsize}
\tablecaption{FF+BB $V$ Light Curves of Ne Novae 2 (Ne2)
\label{light_curves_of_novae_ne2}}
\tablewidth{0pt}
\tablehead{
\colhead{$m_{\rm ff}$} &
\colhead{1.33$M_\sun$} &
\colhead{1.35$M_\sun$} &
\colhead{1.37$M_\sun$} \\
\colhead{(mag)} &
\colhead{(day)} &
\colhead{(day)} &
\colhead{(day)} 
}
\colnumbers
\startdata
  4.500     & 0.0 & 0.0 & 0.0  \\
  4.750     & 0.4617     & 0.6000     & 0.4350     \\
  5.000     & 0.8397     &  1.019     & 0.7630     \\
  5.250     &  1.173     &  1.354     &  1.074     \\
  5.500     &  1.575     &  1.673     &  1.385     \\
  5.750     &  1.992     &  2.019     &  1.694     \\
  6.000     &  2.396     &  2.399     &  2.007     \\
  6.250     &  2.786     &  2.742     &  2.328     \\
  6.500     &  3.189     &  3.075     &  2.628     \\
  6.750     &  3.591     &  3.408     &  2.871     \\
  7.000     &  4.001     &  3.736     &  3.137     \\
  7.250     &  4.403     &  4.041     &  3.380     \\
  7.500     &  4.810     &  4.346     &  3.606     \\
  7.750     &  5.235     &  4.643     &  3.819     \\
  8.000     &  5.717     &  4.959     &  4.044     \\
  8.250     &  6.355     &  5.326     &  4.295     \\
  8.500     &  7.079     &  5.766     &  4.594     \\
  8.750     &  7.949     &  6.299     &  4.943     \\
  9.000     &  8.922     &  6.945     &  5.373     \\
  9.250     &  10.05     &  7.683     &  5.857     \\
  9.500     &  11.45     &  8.533     &  6.443     \\
  9.750     &  13.13     &  9.533     &  7.097     \\
  10.00     &  14.94     &  10.76     &  7.918     \\
  10.25     &  16.83     &  12.15     &  8.828     \\
  10.50     &  18.68     &  13.51     &  9.748     \\
  10.75     &  20.56     &  14.85     &  10.62     \\
  11.00     &  22.30     &  16.25     &  11.40     \\
  11.25     &  23.73     &  17.56     &  12.23     \\
  11.50     &  25.19     &  18.75     &  13.10     \\
  11.75     &  26.75     &  20.01     &  14.03     \\
  12.00     &  28.39     &  21.35     &  15.01     \\
  12.25     &  30.13     &  22.76     &  16.05     \\
  12.50     &  31.97     &  24.26     &  17.15     \\
  12.75     &  33.93     &  25.84     &  18.31     \\
  13.00     &  35.99     &  27.52     &  19.55     \\
  13.25     &  38.18     &  29.30     &  20.86     \\
  13.50     &  40.50     &  31.19     &  22.24     \\
  13.75     &  42.96     &  33.18     &  23.71     \\
  14.00     &  45.57     &  35.30     &  25.27     \\
  14.25     &  48.32     &  37.54     &  26.91     \\
  14.50     &  51.24     &  39.91     &  28.66     \\
  14.75     &  54.34     &  42.42     &  30.50     \\
  15.00     &  57.61     &  45.08     &  32.46     \\
\hline
X-ray
 & 23.4  & 14.3 & 7.80  \\
\hline
$\log f_{\rm s}$
 & $-0.63$  & $-0.75$ & $-0.92$  \\
\hline
$M_{\rm w}$
 & $-0.1$ & $-0.4$  & $-0.8$  \\
\enddata
\end{deluxetable}


\startlongtable
\begin{deluxetable}{llll}
\tabletypesize{\scriptsize}
\tablecaption{FF+BB $V$ Light Curves of Ne Novae 3 (Ne3)
\label{light_curves_of_novae_ne3}}
\tablewidth{0pt}
\tablehead{
\colhead{$m_{\rm ff}$} &
\colhead{1.33$M_\sun$} &
\colhead{1.35$M_\sun$} &
\colhead{1.37$M_\sun$} \\
\colhead{(mag)} &
\colhead{(day)} &
\colhead{(day)} &
\colhead{(day)} 
}
\colnumbers
\startdata
  3.750     & 0.0 & 0.0 & 0.0 \\
  4.000     & 0.4317     & 0.4570     & 0.3855     \\
  4.250     & 0.9427     & 0.9240     & 0.9118     \\
  4.500     &  1.481     &  1.388     &  1.420     \\
  4.750     &  2.023     &  1.914     &  1.889     \\
  5.000     &  2.594     &  2.477     &  2.324     \\
  5.250     &  3.173     &  3.008     &  2.752     \\
  5.500     &  3.686     &  3.426     &  3.140     \\
  5.750     &  4.176     &  3.861     &  3.498     \\
  6.000     &  4.665     &  4.270     &  3.870     \\
  6.250     &  5.128     &  4.653     &  4.180     \\
  6.500     &  5.621     &  5.042     &  4.487     \\
  6.750     &  6.130     &  5.430     &  4.798     \\
  7.000     &  6.622     &  5.842     &  5.058     \\
  7.250     &  7.129     &  6.232     &  5.320     \\
  7.500     &  7.646     &  6.621     &  5.605     \\
  7.750     &  8.221     &  7.041     &  5.930     \\
  8.000     &  8.886     &  7.557     &  6.296     \\
  8.250     &  9.698     &  8.153     &  6.707     \\
  8.500     &  10.63     &  8.908     &  7.222     \\
  8.750     &  11.80     &  9.778     &  7.801     \\
  9.000     &  13.14     &  10.79     &  8.494     \\
  9.250     &  14.64     &  11.92     &  9.268     \\
  9.500     &  16.36     &  13.22     &  10.21     \\
  9.750     &  18.38     &  14.81     &  11.29     \\
  10.00     &  20.61     &  16.55     &  12.48     \\
  10.25     &  23.04     &  18.36     &  13.69     \\
  10.50     &  25.45     &  20.14     &  14.96     \\
  10.75     &  27.66     &  21.95     &  16.09     \\
  11.00     &  30.02     &  23.34     &  17.07     \\
  11.25     &  32.48     &  24.82     &  18.12     \\
  11.50     &  34.55     &  26.38     &  19.23     \\
  11.75     &  36.64     &  28.03     &  20.40     \\
  12.00     &  38.87     &  29.78     &  21.64     \\
  12.25     &  41.22     &  31.64     &  22.95     \\
  12.50     &  43.71     &  33.60     &  24.35     \\
  12.75     &  46.35     &  35.69     &  25.83     \\
  13.00     &  49.15     &  37.89     &  27.39     \\
  13.25     &  52.10     &  40.23     &  29.04     \\
  13.50     &  55.24     &  42.70     &  30.80     \\
  13.75     &  58.56     &  45.32     &  32.66     \\
  14.00     &  62.08     &  48.10     &  34.62     \\
  14.25     &  65.81     &  51.04     &  36.71     \\
  14.50     &  69.76     &  54.16     &  38.92     \\
  14.75     &  73.94     &  57.46     &  41.26     \\
  15.00     &  78.37     &  60.95     &  43.73     \\
\hline
X-ray
 & 41.6  & 24.1 & 12.4  \\
\hline
$\log f_{\rm s}$
 & $-0.51$  & $-0.63$ & $-0.80$  \\
\hline
$M_{\rm w}$
 & $0.3$ & $0.0$  & $-0.4$  \\
\enddata
\end{deluxetable}



\begin{figure}
\gridline{\fig{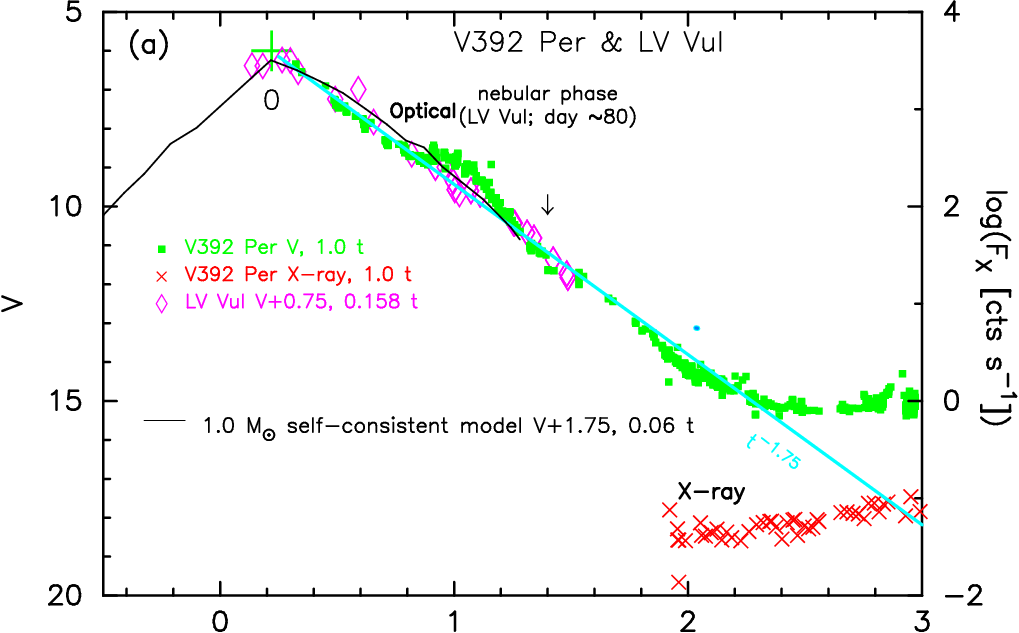}{0.45\textwidth}{}
          }
\gridline{
          \fig{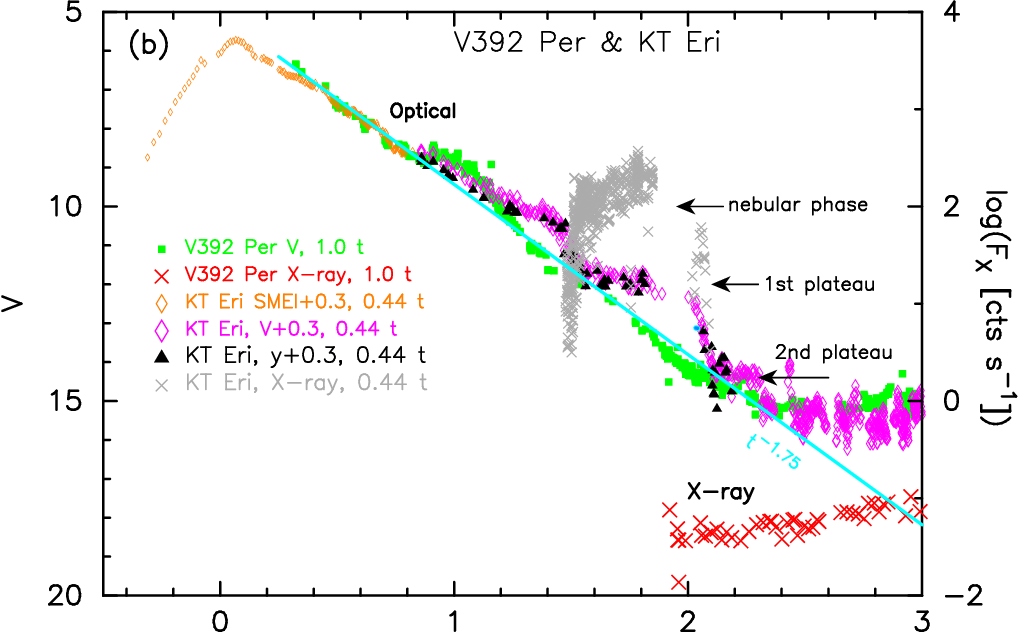}{0.45\textwidth}{}
          }
\gridline{
          \fig{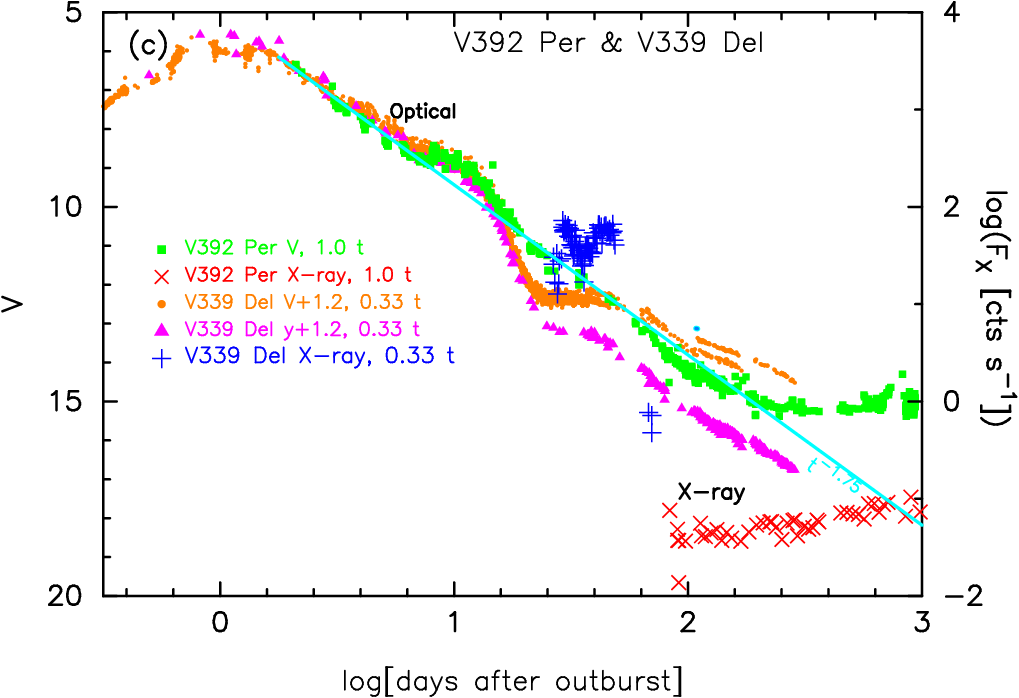}{0.45\textwidth}{}
          }
\caption{
(a) Two $V$ light curves of V392 Per and LV Vul are
overlapped along Equation (\ref{overlap_brigheness}).
The text ``LV Vul V+0.75, 0.158 t'', for example, means $f_{\rm s}=0.158$ and
$\Delta V=+0.75$, for the template nova LV Vul against the $V$ light curve
of the target nova V392 Per (``V392 Per V, 1.0 t'').  
We also add \citet{kat22sha}'s fully self-consistent nova model (thin black
line). The $V$ peak (green plus symbol labeled ``0'') of V392 Per 
is overlapped with the peak of this self-consistent nova model. 
(b) Same as panel (a), but for the template nova KT Eri.
(c) Same as panel (a), but for the template nova V339 Del.
The data of LV Vul, KT Eri, and V339 Del are the same as those
in \citet{hac23k}, \citet{hac25kw}, and \citet{hac24km}, respectively. 
\label{v392_per_kt_eri_v339_del_v_x_stretching}}
\end{figure}

\begin{figure}
\epsscale{1.0}
\plotone{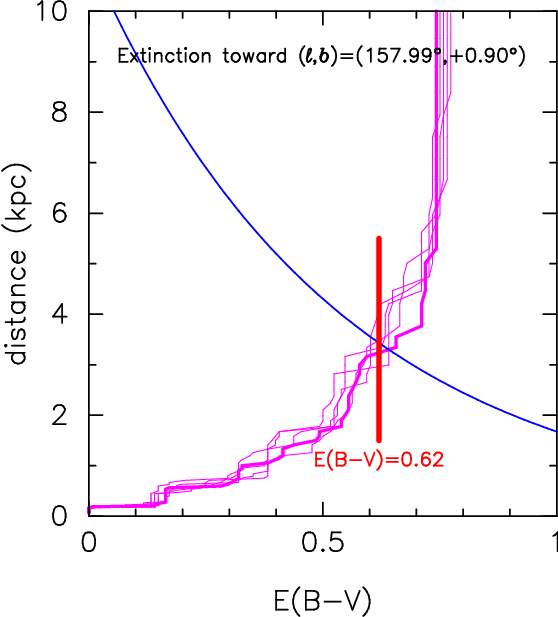}
\caption{
The distance-reddening relations toward V392 Per whose galactic coordinates
are $(\ell, b)= (157\fdg99, +0\fdg90)$.
The blue line denotes the relation of Equation (\ref{distance-reddening_law})
together with $(m-M)_V=14.6$ for V392 Per.  The thin magenta lines are
the sample distance-reddening relations given by \citet{gre19}
while the thick magenta line is their best-fit line for them.
Here, we use the relation of $E(B-V)= 0.884\times$(Bayestar19) \citep{gre19}.
The two relations (blue and magenta) cross at the distance of 
$d\approx 3.5$ kpc and $E(B-V)\approx 0.62$.
\label{distance_reddening_relation_v392_per}}
\end{figure}

\section{Time-stretching Method}
\label{time_stretching_method}

To analyze the light curve of a nova, its distance 
is one of the most important parameters.  The time-stretching method   
\citep{hac10k, hac15k, hac16k, hac18k,hac20skhs} is a useful way 
to derive the distance modulus in the $V$ band, $(m-M)_V$.
In this appendix, we explain the time-stretching method and
determine the distance to V392 Per.

This method is based on the similarity between two nova light curves.
Adopting an appropriate time-stretching parameter, we are able to overlap
two nova light curves even if the two nova speed classes are different. 
If the two nova $V$ light curves, i.e.,
one is called the template and the other is called the target,
$(m[t])_{V,\rm target}$ and $(m[t])_{V,\rm template}$
overlap each other after time-stretching by a factor of $f_{\rm s}$
in the horizontal direction and shifting vertically down by $\Delta V$, i.e.,
\begin{equation}
(m[t])_{V,\rm target} = \left((m[t \times f_{\rm s}])_V
+ \Delta V\right)_{\rm template},
\label{overlap_brigheness}
\end{equation}
their distance moduli in the $V$ band satisfy
\begin{eqnarray}
& & (m-M)_{V,\rm target} \cr
&=& ( (m-M)_V + \Delta V )_{\rm template} - 2.5 \log f_{\rm s}.
\label{distance_modulus_formula}
\end{eqnarray}
Here, $m_V$ and $M_V$ are the apparent and absolute $V$ magnitudes,
and $(m-M)_{V, \rm target}$ and $(m-M)_{V, \rm template}$ are
the distance moduli in the $V$ band
of the target and template novae, respectively.
\citet{hac18k, hac18kb, hac19k, hac19kb, hac21k} confirmed that
Equations (\ref{overlap_brigheness}) and (\ref{distance_modulus_formula})
are also broadly valid for other $U$, $B$, and $I$ (or $I_{\rm C}$) bands.

\subsection{Distance Modulus in the $V$ band}
\label{distance_v_band}

This remarkable similarity is demonstrated in
Figure \ref{v392_per_kt_eri_v339_del_v_x_stretching}, which
shows the $V$ and X-ray light curves for (a) V392 Per and LV Vul,
(b) V392 Per and KT Eri, (c) V392 Per and V339 Del.
These novae have rather different timescales of $V$ light curve declines
that are time-stretched into almost one line in the figure. 

In Figure \ref{v392_per_kt_eri_v339_del_v_x_stretching}a,
we regard V392 Per as the target and LV Vul as the template
in Equation (\ref{overlap_brigheness}).  We adopt $f_{\rm s}= 0.158$
and $\Delta V= +0.75$ and have the relation of 
\begin{eqnarray}
(m&-&M)_{V, \rm V392~Per} \cr
& = & (m - M + \Delta V)_{V, \rm LV~Vul} - 2.5 \log 0.158 \cr
&=& 11.85 + 0.75\pm 0.2 + 2.0 = 14.6\pm 0.2,
\label{distance_modulus_v392_per_lv_vul_v}
\end{eqnarray}
where we adopt $(m-M)_{V, \rm LV~Vul}=11.85$ from \citet{hac18k}.

We similarly apply our time-stretching method to a pair of V392 Per and
KT Eri, and a pair of V392 Per and V339 Del,
and plot them in Figure \ref{v392_per_kt_eri_v339_del_v_x_stretching}b
and c, respectively.
Then, we have
\begin{eqnarray}
(m&-&M)_{V, \rm V392~Per} \cr
&=& ((m - M)_V + \Delta V)_{\rm KT~Eri} - 2.5 \log 0.44 \cr
&=& 13.4 + 0.3\pm0.2 + 0.9 = 14.6\pm0.2 \cr
&=& ((m - M)_V + \Delta V)_{\rm V339~Del} - 2.5 \log 0.33 \cr
&=& 12.2 + 1.2\pm0.2 + 1.2 = 14.6\pm0.2,
\label{distance_modulus_v_v1535_sco_v339_del_kt_eri}
\end{eqnarray}
where we adopt $(m-M)_{V, \rm KT~Eri}=13.4$ from \citet{hac25kw} and
$(m-M)_{V, \rm V339~Del}= 12.2$ from \citet{hac24km}.
Thus, we obtain $(m-M)_{V, \rm V392~Per}= 14.6\pm 0.2$, 
which is consistent with the result in Section
\ref{distance_reddening_orbit}.


\begin{figure*}
\gridline{\fig{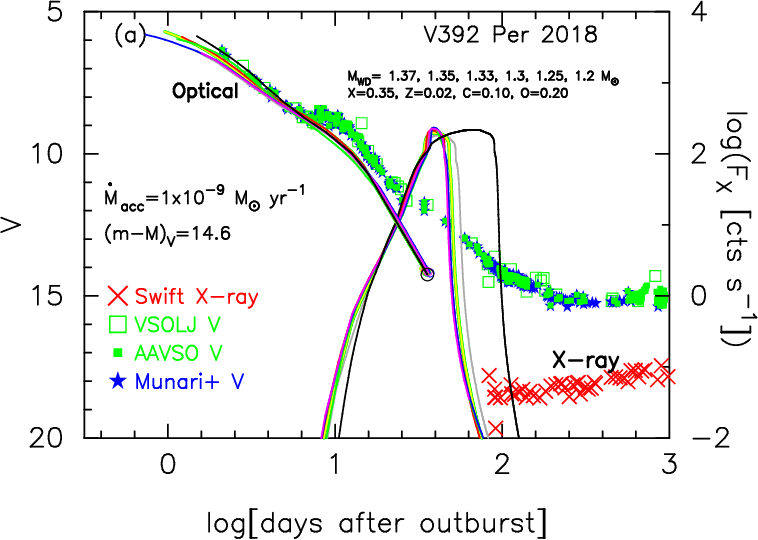}{0.45\textwidth}{}
          \fig{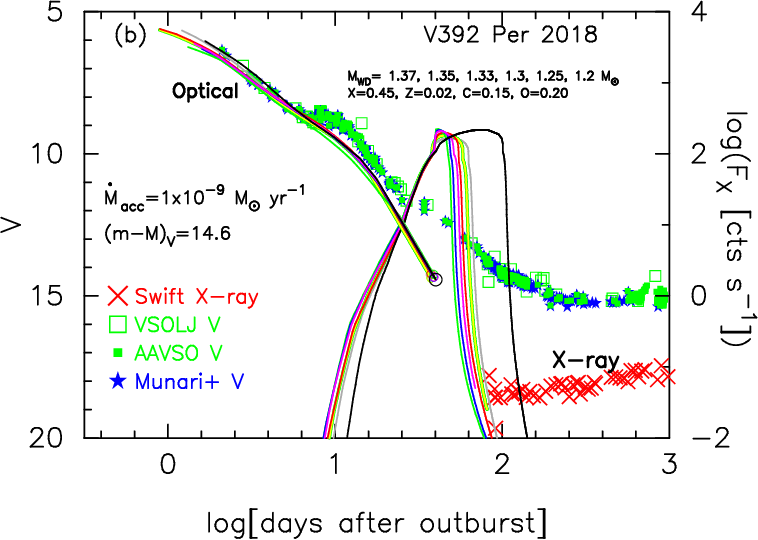}{0.45\textwidth}{}
          }
\gridline{
          \fig{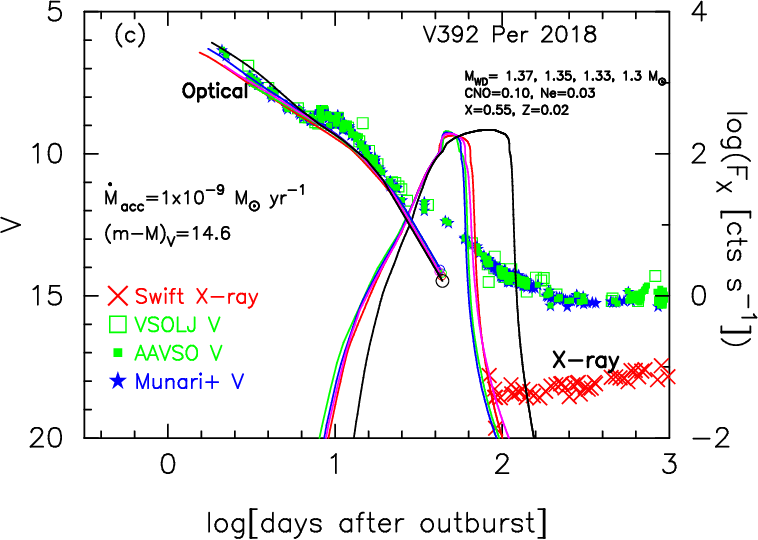}{0.45\textwidth}{}
          \fig{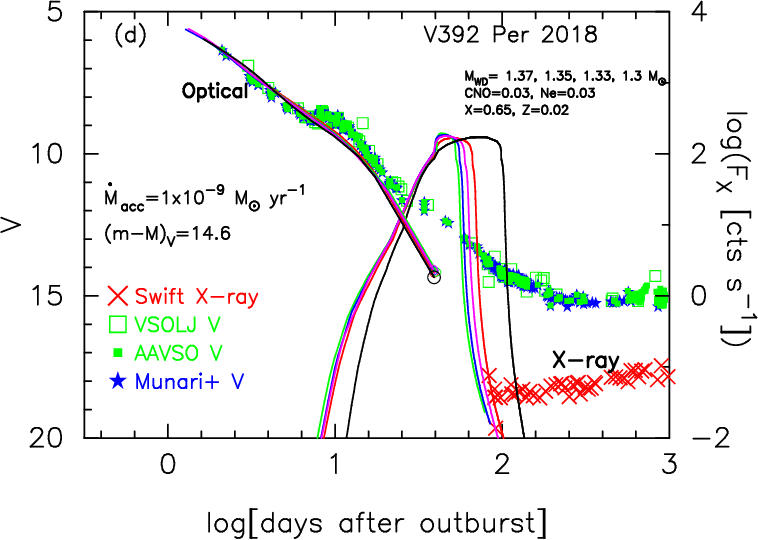}{0.45\textwidth}{}
          }

\caption{
Same as those in Figure \ref{individual_v392_per_v_chemical_logscale_no2},
but each FF+BB $V$ and model X-ray light curves are time-stretched with the 
stretching factor $\log f_{\rm s}$ in 
(a) Table \ref{light_curves_of_novae_co2},
(b) Table \ref{light_curves_of_novae_co3},
(c) Table \ref{light_curves_of_novae_ne2}, and
(d) Table \ref{light_curves_of_novae_ne3}.
We add a $0.98 ~M_\sun$ (CO3) WD model to each panel by the solid black line.
\label{unify_v1535_sco_v_chemical_logscale_no2}}
\end{figure*}


\begin{figure*}
\gridline{\fig{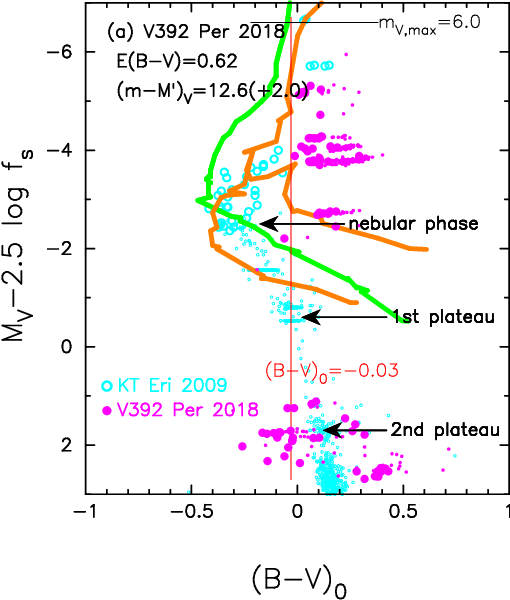}{0.45\textwidth}{}
          \fig{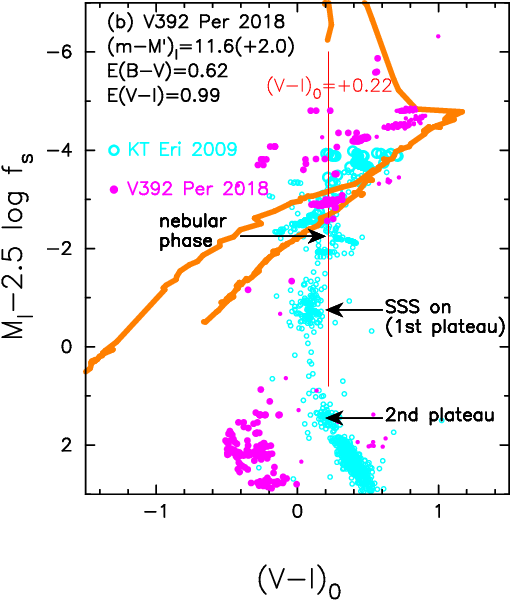}{0.45\textwidth}{}
          }
\caption{
(a) The time-stretched $(B-V)_0$-$(M_V-2.5 \log f_{\rm s})$ color-magnitude
diagram for V392 Per (filled magenta circles) and KT Eri (open cyan circles).
The data of V392 Per are taken from \citet{mun20mm} (large filled magenta
circles), AAVSO (small magenta circles), and VSOLJ (small magenta circles).
The data of KT Eri are taken from \citet{ima12t} and SMARTS \citep{wal12bt}.
Each start of the nebular, first (1st) plateau, and second (2nd) plateau
phases of KT Eri are indicated by arrows, which are shown in Figure
\ref{v392_per_kt_eri_v339_del_v_x_stretching}b, taken from \citet{hac25kw}.
The vertical solid red line of $(B-V)_0= -0.03$ is the
intrinsic color of optically thick free-free emission \citep{hac14k}.
The solid green and orange lines denote the template tracks
of V1500~Cyg and LV~Vul, respectively.
The track of LV~Vul splits into two branches in the later
phase.  See \citet{hac16kb} and \citet{hac19k}
for details of these template tracks.
(b) The time-stretched $(V-I)_0$-$(M_I-2.5 \log f_{\rm s})$ color-magnitude
diagram for V392 Per and KT Eri.
The data of V392 Per and KT Eri are taken from the same sources
as in panel (a).  The thick solid lines
of orange correspond to the outburst track of LV Vul,
which are taken from \citet{hac21k}.
The vertical solid red line of $(V-I)_0= +0.22$ is the
intrinsic color of optically thick free-free emission \citep{hac21k}.
See Figure 12 of \citet{hac25kw} for more details on the KT Eri track.
\label{hr_diagram_v392_per_kt_eri_bv_vi}
}
\end{figure*}

\subsection{Outburst day from time-stretched light curves}
\label{outburst_day_from_time-streching}

Figure \ref{v392_per_kt_eri_v339_del_v_x_stretching}a also show the 
time-stretched FF+BB light curve \citep[black line;][]{hac23k}
based on \citet{kat22sha}'s fully self-consistent nova model.
The $V$ peaks of LV Vul and V392 Per (green plus symbol labeled ``0'') are
well overlapped.  Therefore, we are able to estimate the outburst day
(when thermonuclear runaway starts) from this self-consistent nova model
with the time-stretching factor of $f_{\rm s}= 0.06$ against the V392 Per
$V$ light curve.  The rising time to the peak is about
$(\Delta t)_{\rm rise}\approx 10^{0.22} =$ 1.66 days.
We adopt the outburst day to be $t_{\rm OB}=$ JD 2,458,236.2.

\subsection{Distance-reddening relation toward V392 Per}
\label{distance-reddening_relation_v392_per}

A three-dimensional absorption map (Bayestar19) in our Galaxy was given by 
\citet{gre19}.  We plot its distance-reddening relation (thin magenta
lines) toward V392 Per in Figure \ref{distance_reddening_relation_v392_per},
where we use the relation of $E(B-V)= 0.884\times$(Bayestar19).
The thick magenta line is a best-fit one among them.
We overplot the relation (blue line) 
of Equation (\ref{distance-reddening_law}) 
together with our obtained value of
$(m-M)_V=14.6$ from the time-stretching method.

These two (blue and magenta) relations cross at 
the distance of $d\approx 3.5$ kpc
and the reddening of $E(B-V)\approx 0.62$
\citep[see also ][]{mun20mm}.
This distance is consistent with the result of Gaia eDR3
\citep[$d=3.45^{+0.62}_{-0.51}$ kpc,][]{bai21rf}. 
Green et al.'s distance-reddening relation gives a value of
$E(B-V)= 0.884 \times$(Bayestar19) $= 0.884\times 0.71^{+0.03}_{-0.02}$
$= 0.62^{+0.03}_{-0.02}$ at the distance of $d=3.45$ kpc.

\subsection{Time-stretching model light curves}
\label{time-streching_model_light_curve}

Our FF+BB light curve models also satisfy the time-stretching
relations defined by Equations (\ref{overlap_brigheness})
and (\ref{distance_modulus_formula}).
We time-stretch each FF+BB light curve in Figure
\ref{individual_v392_per_v_chemical_logscale_no2}a 
and replot them in Figure 
\ref{unify_v1535_sco_v_chemical_logscale_no2}a.
All the CO2 models overlap on the $1.33 ~M_\sun$ WD
with each time-stretching factor.
To demonstrate the similarity we added
the $0.98 ~M_\sun$ (CO3) WD model (black line), which is 
a well fitted model to the LV Vul $V$ light curve.
We use the time-stretching factor of $\log f_{\rm s}$  in
Table \ref{light_curves_of_novae_co2} all against that of LV Vul.

In Figure \ref{unify_v1535_sco_v_chemical_logscale_no2}b, 
we similarly overplot all the CO3 models on the $1.33 ~M_\sun$ WD model
as well as the $0.98 ~M_\sun$ (CO3) WD model.
Here, we use the time-stretching factor of $\log f_{\rm s}$  in
Table \ref{light_curves_of_novae_co3}.

Figure \ref{unify_v1535_sco_v_chemical_logscale_no2}c shows
all the Ne2 models overlapped on the $1.35 ~M_\sun$ WD model
as well as the $0.98 ~M_\sun$ (CO3) WD model.
Here, we use the time-stretching factor of $\log f_{\rm s}$  in
Table \ref{light_curves_of_novae_ne2}.

The Ne3 models are all overlapped on the $1.37 ~M_\sun$ WD model
as well as the $0.98 ~M_\sun$ (CO3) WD model.
Here, we use the time-stretching factor of $\log f_{\rm s}$  in
Table \ref{light_curves_of_novae_ne3}.

\subsection{Color-magnitude diagrams of V392 Per and KT Eri}
\label{color-magnitude_diagram}

When the $B$ and $V$ light curves of the target nova overlap with
the $B$ and $V$ light curves of the template nova, respectively, by
the same time-stretching method, i.e., by the same time-stretching factor
of $f_{\rm s}$, the intrinsic $(B-V)_0$ color curve of the target nova
also overlaps with the intrinsic $(B-V)_0$ color curve of the template nova.
This means that the time-stretched $(B-V)_0$-$(M_V-2.5 \log f_{\rm s})$
color-magnitude diagrams of the target and template novae
overlap with each other \citep{hac19k}.

Figure \ref{hr_diagram_v392_per_kt_eri_bv_vi}a shows the
$(B-V)_0$-$(M_V-2.5 \log f_{\rm s})$ diagram for V392 Per and KT Eri.
In the figures, we adopt $f_{\rm s}= 0.158$ for V392 Per and
$f_{\rm s}= 0.36$ for KT Eri both against LV Vul.  The text of
``$(m-M')_V= 12.6(+2.0)$'' in Figure \ref{hr_diagram_v392_per_kt_eri_bv_vi}a
means the time-stretching distance modulus in the $V$ band, that is,
$(m-M')_V\equiv (m-(M - 2.5 \log f_{\rm s}))_V = 12.6$ and
$(+2.0) \equiv - 2.5 \log f_{\rm s} = -2.5 \log 0.158 = +2.0$ for V392 Per.
Then, the distance modulus in the $V$ band is $(m-M)_V= 12.6 + 2.0 = 14.6$.
The data of V392 Per are taken from \citet{mun20mm}, AAVSO, and VSOLJ
while the data of KT Eri are from SMARTS \citep{wal12bt} and \citet{ima12t}.
We also plot the time-stretched color-magnitude diagrams for
LV Vul (orange line) and V1500 Cyg (green line).
All the data of these two novae are the same as those in \citet{hac19k}.

There is a large difference between a pair of V392 Per and KT Eri and
other classical novae LV Vul and V1500 Cyg:
the trend of $(B-V)_0$ color in the later phase of LV Vul and V1500 Cyg
is toward red, but that of KT Eri
seems to stay at/around $(B-V)_0\sim 0.0$ in the first plateau
and then $(B-V)_0\sim 0.2$ in the second plateau of KT Eri (see Figures
\ref{V392_per_kt_eri_v339_del_bvi_logscale}b and
\ref{v392_per_kt_eri_v339_del_v_x_stretching}b, for first plateau
and second plateau of KT Eri).
The color of $(B-V)_0\sim 0.0$ is a typical one for irradiated
accretion disks as frequently observed in recurrent novae like in U Sco
\citep[see Figure 29(d) of ][]{hac21k}.
The $(B-V)_0$ color of V392 Per also stays at $(B-V)_0\sim 0.0$
in the viscous heating disk phase (the same as the second plateau in KT Eri).
Thus, we interpret that the color $(B-V)_0\sim 0.0$ (or $(B-V)_0\sim 0.2$)
in the later phase at $M_V\sim 0.0$ (or $m_V\equiv V\sim 14.6$ for V392 Per)
is due to a large optical contribution from the accretion disk.    

We also plot the $(V-I)_0$-$(M_I-2.5 \log f_{\rm s})$ diagram for V392 Per
and KT Eri in Figure \ref{hr_diagram_v392_per_kt_eri_bv_vi}b.
Here, $I$ corresponds to $I_{\rm C}$.  The $(V-I)_0$ color
of KT Eri is close to $0.0$-$0.2$ in the later phase of the outburst
because the disk dominates the optical flux of KT Eri.  The track of
V392 Per in the later phase is similar to that of, but located 
at a bluer side $(V-I)_0\sim -0.3$ of, KT Eri.  The details of
each tracks of the other novae were discussed in \citet{hac21k}.

\end{document}